\documentclass[12pt, a4paper]{article}
\pdfoutput=1
\usepackage[utf8]{inputenc}
\usepackage{fancybox}
\usepackage[T1]{fontenc}
\usepackage{lmodern, textcomp}
\usepackage{bm}
\usepackage[american]{babel}
\usepackage{csquotes}

\usepackage{eurosym}
\usepackage[nottoc]{tocbibind}
\usepackage{authblk}
\usepackage{fancyhdr}
\usepackage{xspace}

\usepackage{graphicx}
\usepackage{subcaption}
\usepackage{float}
\usepackage{makecell}
\usepackage{multirow}
\usepackage{multicol} 
\usepackage{marginnote}
\usepackage[multiple]{footmisc}

\usepackage[usenames, dvipsnames, svgnames, table]{xcolor}

\usepackage[backend=bibtex8, citestyle=numeric-comp, maxbibnames=99,
			sorting=none, sortcase=false, sortcites=true, giveninits=true]{biblatex}

\usepackage{amsmath, amsfonts, amssymb}
\usepackage{mathtools}
\usepackage{mathrsfs}
\usepackage{cancel}
\usepackage{braket}
\usepackage{tensor}
\usepackage{slashed}
\usepackage{siunitx}
\usepackage{listings}

\usepackage{stmaryrd}
\usepackage{caption}
\usepackage{relsize,exscale}
\usepackage{array}
\usepackage[toc,page]{appendix}

\usepackage{enumerate}

\DeclareSymbolFont{largesymbols}{OMX}{cmex}{m}{n}

\setcellgapes{1pt}
\makegapedcells
\newcolumntype{R}[1]{>{\raggedleft\arraybackslash }b{#1}}
\newcolumntype{L}[1]{>{\raggedright\arraybackslash }b{#1}}
\newcolumntype{C}[1]{>{\centering\arraybackslash }b{#1}}

\newcommand{\Tr}{\mathrm{Tr}}

\newtheorem{definition}{Definition}

\newtheorem{proposition}{Proposition}
\newtheorem{remark}{Remark}
\newtheorem{claim}{Claim}

\newcommand{\beq}{\begin{equation}}
\newcommand{\eeq}{\end{equation}}
\newcommand{\bea}{\begin{eqnarray}}
\newcommand{\eea}{\end{eqnarray}}
\definecolor{mygray}{gray}{0.3}

\newcommand{\bes}{\begin{eqnarray}}
\newcommand{\ees}{\end{eqnarray}}

\newcommand\restr[2]{{
  \left.\kern-\nulldelimiterspace 
  #1 
  \vphantom{\big|} 
  \right|_{#2} 
  }}
\def\extd{\mathrm {d}}

\usepackage{multicol}
\usepackage{amssymb}

\usepackage[heightrounded, top=3.5cm, bottom=3.5cm, left=3cm, right=3cm, headheight=14pt]{geometry}

\usepackage[pdftex, breaklinks=true, linktocpage,
	colorlinks=true, urlcolor=blue, linkcolor=blue, citecolor=red
]{hyperref}

\newcommand{\email}[1]{\href{mailto:#1}{\nolinkurl{#1}}}
\newcommand{\emailfoot}[1]{\thanks{\email{#1}}}

\usepackage{marginnote}
\newcounter{draftcommentcnt}
\NewDocumentCommand{\draftcomment}{s O{red} m}{%
	\def\margnote{\IfBooleanTF{#1}{\marginnote}{\marginpar}}%
	\stepcounter{draftcommentcnt}%
	\textcolor{#2}{#3}%
	\margnote{\textcolor{#2}{$\Leftarrow$ \arabic{draftcommentcnt}}}%
}

\fancypagestyle{preprint}{
	\fancyhf{}

	\fancyhead[R]{\textsf{\preprint}}
}

\numberwithin{equation}{section}

\hypersetup{
	pdftitle={Stochastic dynamics for Group Field Theories 2},
}

\title{Stochastic dynamics for group field theories II: Methods  for nonequilibrium renormalization group}

\author[1]{Vincent Lahoche\emailfoot{vincent.lahoche@cea.fr}}
\author[2]{Dine Ousmane Samary\emailfoot{dine.ousmanesamary@cipma.uac.bj}}

\affil[1]{%
	Université Paris Saclay, \textsc{Cea}, Gif-sur-Yvette, F-91191, France
}

\affil[2]{%
	Faculté des Sciences et Techniques (ICMPA-UNESCO Chair)
	\protect\\
	Université d'Abomey-Calavi, 072 BP 50, Bénin
}

\begin{document}
\maketitle

\begin{abstract}

This paper is a continuation of our earlier work, which aimed to develop methods for understanding the renormalization group of tensorial group field theories within the stochastic quantization framework. In that first study, we showed that the equations governing melonic structures, together with Ward identities, make it possible to close the hierarchy of flow equations, thereby reproducing the results of equilibrium theory. In the present work, we go further by extending the formalism to the out-of-equilibrium regime, while also examining the stability of dynamical equilibrium, specifically, potential violations of the fluctuation–dissipation theorem. Our objective here is purely methodological, and we focus on a simplified “toy” Abelian model that retains only the characteristic non-localities of group field theories.


\end{abstract}
\pagebreak

\tableofcontents

\section{Introduction}



Nowadays, gravity remains the last fundamental interaction resisting quantization, while the other three interactions have successfully undergone quantization (albeit with subtleties, particularly for the strong interaction) \cite{weinberg1995quantum}. Several approaches to quantum gravity have been proposed, the most prominent, judging by the number of adherents, being string theory \cite{zwiebach2004first} and loop quantum gravity (LQG) \cite{rovelli2004quantum}. The latter belongs to a broader class of so-called discrete approaches, with dynamical triangulations providing one alternative example \cite{Loll_2019}. Group field theory (GFT) has emerged as a promising framework for addressing the problem of LQG dynamics within a field-theoretic setting, where Feynman diagrams naturally take the form of spin foams (see \cite{Freidel_2005,baratin2012ten} or \cite{oriti2021tensorial,gielen2025hilbert} for recent results, including developments in quantum cosmology and Hilbert space formulations).

A crucial contribution to the development of GFTs came from random matrix models \cite{Francesco_1995}, and their higher-dimensional generalizations, the colored tensor models \cite{Gurau_2016,gurau2017random}. Like matrix models, these theories have the advantage of being characterized by a well-defined power counting. This property is essential for establishing a robust renormalization program and led to the formulation of tensorial group field theories (TGFTs) \cite{carrozza2015discrete,carrozza2014renormalization2,carrozza2014renormalization,Carrozza_2016ccc}. For these theories, solid renormalization theorems have been established, along with non-perturbative analyses \cite{Lahoche_2020b,Lahoche_2020d,Lahoche_2021c,Carrozza_2017,Carrozza_2017a,Geloun_2016,Ben_Geloun_2015,Geloun_2018,marchetti2021phase,pithis2021no}. Moreover, the role of phase transitions in the geometrogenesis scenario highlights the importance of studying the renormalization group in GFTs. Progress has been made in this direction since the introduction of TGFTs more than a decade ago, but the main difficulty lies in the specific non-local structure of their interactions, which introduces important novelties compared to ordinary field theories. In particular, the fact that the anomalous dimension already plays a role in the symmetric phase raises subtleties that cast doubt on the reliability of common approximations \cite{Lahoche_2019bb}.

This article continues within this framework and builds on our previous work \cite{lahoche2023stochastic}, from which we borrow several results. As in that earlier study, our objective is not to solve a specific problem in quantum gravity but rather to provide methodological tools. To this end, we consider a “toy” model chosen to be as simple as possible: an Abelian theory known to be just-renormalizable \cite{carrozza2014renormalization}. The model we study is stochastic, described by a Langevin-type tensor equation, and we analyze it using the functional renormalization group. In our first paper, we restricted attention to equilibrium dynamics, an assumption that we relax in this work. We show how, in the symmetric phase, the structure of melonic interactions dominating the UV flow, together with the non-trivial Ward identities of the model (arising because unitary invariance is broken by the classical action), makes it possible to close the hierarchy of flow equations. This strategy has already proven effective: the same authors developed it for equilibrium GFTs in \cite{Lahoche_2019bb,Lahoche:2018oeo}, where its advantages compared to the standard vertex expansion were discussed in detail.

We will not revisit the physical foundations of the model, which were already presented in detail in \cite{lahoche2023stochastic}\footnote{While preparing this paper, we discovered some typos in the computations of that reference, which invalidate certain conclusions but not the overall methodology. Readers are encouraged to consult the revised arXiv version.}. Let us briefly summarize the essential points. At the mathematical level, our approach can be viewed as a stochastic quantization of a GFT. Physically, “time” may be interpreted relationally as the state of a scalar field weakly coupled to the quantum gravitational field\footnote{In this setting, the GFT toy model plays the role of gravitational degrees of freedom.}, which serves as a physical clock (in a regime where quantum effects of the scalar field itself can be neglected). This relational perspective can be regarded as a semi-quantum counterpart of the classical view in general relativity, where contiguity relations between fields define spacetime properties, and coupling to gravitation defines the metric field \cite{rovelli_2004,connes1994neumann,https://doi.org/10.48550/arxiv.1807.04875,https://doi.org/10.48550/arxiv.2110.08641,marchetti2021effective,wilson2019relational,li2017group,https://doi.org/10.48550/arxiv.2112.02585,marchetti2021phase}\\

\section{Model and conventions}

In this section, we define the stochastic models, the path integral representation and the conventions used in the rest of the paper. For a general review on this topic, the reader may consult the standard references \cite{livi2017nonequilibrium,Zinn-Justin:1989rgp,ZinnJustinBook2}. \\

To put in a nutshell, a group field $\varphi$ is a field defined on $d$-copies of some group manifold $\bm{\mathrm{G}}$:
\begin{equation}
\varphi: (g_1,\cdots, g_d)\in (\bm{\mathrm{G}})^{\times d} \to \varphi(g_1,\cdots,g_d)\in \mathbb{K} \,.
\end{equation}
We focus on complex group fields, i.e., $\mathbb{K} \equiv \mathbb{C}$. For convenience, we denote by $\bm{g} := (g_1, \cdots, g_d)$ the elements of $(\bm{\mathrm{G}})^{\times d}$, and by $\varphi(\bm{g})$ the value of the field at the point $\bm{g} \in (\bm{\mathrm{G}})^{\times d}$. We generally assume that the classical field $\varphi$ is square-integrable, with the standard $L^2((\bm{\mathrm{G}})^{\times d})$ norm defined by the inner product:
\begin{equation}
(\varphi, \varphi^\prime):= \int d\bm{g} \,\bar{\varphi}(\bm g) \varphi^\prime(\bm g)
\end{equation}
is assumed to be bounded: $\Vert \varphi \Vert := (\varphi, \varphi) < \infty$. Here, $\bar{\varphi}$ is the standard complex conjugation of $\varphi$ and $d\bm g$ means
\begin{equation}
d\bm g:= dg_1dg_2\cdots dg_d\,,
\end{equation}
where $dg_\ell$ is the Haar measure over $\textbf{G}$, normalized such that:
\begin{equation}
\int d\bm g =1\,.
\end{equation}
For the purpose of this paper, we further assume that this field is a dynamic variable, depending on time $t$, which is taken to be a real variable. Finally, the evolution of the field is postulated to satisfy the dissipative Langevin equation:
\begin{equation}
\boxed{
\dot{{\varphi}}(\bm g,t)=- \Omega \frac{\partial }{\partial \bar{\varphi}(\bm g,t)} \mathcal{H}[\varphi,\bar{\varphi}]+\eta(\bm g,t)\,,}\label{langevin}
\end{equation}
where the noise $\eta(\bm g,t)$ is a Gaussian random group field with zero mean and variance:
\begin{equation}
\langle \eta(\bm g, t) \bar{\eta}(\bm g^\prime, t^\prime) \rangle_\eta =\Omega\, \delta(\bm g^\prime (\bm g)^{-1}) \delta(t-t^\prime)\,,\label{noisecorrelation}
\end{equation}
the notation $\langle X \rangle_\eta$ meaning average over $\eta$ with respect to some Gaussian probability density $d\rho(\eta)/d[\eta]$. Moreover $\delta(\bm g^\prime (\bm g)^{-1}) := \prod_{\ell=1}^d \delta(g_\ell^\prime g_\ell^{-1})\,,$ the Dirac delta $\delta(g_\ell g_\ell^{-1})$ being such that:
\begin{equation}
\int dg \, \delta(g^\prime (g)^{-1}) f(g)=f(g^\prime)\,.
\end{equation}
Finally, $\Omega > 0$ denotes an arbitrary time scale, the “dot” stands for $d/dt$, and the \textit{classical Hamiltonian} $\mathcal{H}$ is assumed to be invariant under unitary transformations. Occasionally, we will also use the dot to indicate the RG flow, but the context should prevent any ambiguity. \\

The behavior of the stochastic field can be understood in a probabilistic description. Let us denote as $\bm q(t):=\{ \varphi(\bm g,t), \bar{\varphi}(\bm g,t)\}$ a given "position" for the random complex field in the configuration space. For some initial conditions $\bm{q}(t^\prime)=\bm{q}^\prime$ fixed at time $t^\prime$, the transition probability $P(\bm q, t;\bm q^\prime, t^\prime)$ to find the system at the place $\bm q$ at time $t>t^\prime$ is:
\begin{equation}
P(\bm q, t;\bm q^\prime, t^\prime)=\big\langle \delta(\bm q(t)-\bm q) \big\rangle_\eta\,. \label{eqProba}
\end{equation}
Note that it is convenient to assume initial conditions and to adopt the simplified notation $P(\bm{q}, t)$ for $P(\bm{q}, t; \bm{q}^\prime, t^\prime)$. The transition probability follows a Markov process and satisfies a Fokker–Planck–type equation, which can be derived from \eqref{langevin} (see, for instance, \cite{ZinnJustinBook2}). Explicitly:
\begin{equation}
\frac{\partial }{\partial t} P(\bm q,t)=\Omega \hat{H} P(\bm q,t)\,,
\end{equation}
with:
\begin{equation}
\hat{H}:=\int d\bm g \left(\frac{\delta^2 }{\delta \varphi(\bm g)\partial \bar{\varphi}(\bm g)}+2\frac{\delta^2 \mathcal{H}}{\delta \varphi(\bm g)\partial \bar{\varphi}(\bm g)} +\frac{\delta \mathcal{H}}{\delta \varphi(\bm g)}\frac{\delta}{\delta \bar{\varphi}(\bm g)}+\frac{\delta \mathcal{H}}{\delta \bar{\varphi}(\bm g)}\frac{\delta}{\delta \varphi(\bm g)}\right)\,.
\end{equation}
The long time solution,
\begin{equation}
\rho(\bm q):=\lim\limits_{t \to +\infty} P(\bm q,t;\bm {q}^\prime, t^\prime)\,,
\end{equation}
which corresponds to stationary solutions of the Fokker-Planck equation if the system reach equilibrium, is given by: 
\begin{equation}
\rho(\bm q)\propto e^{-2\mathcal{H}}\,,
\end{equation}
and exists if it is normalizable, i.e., if the corresponding partition function exists. In the equilibrium dynamics regime, the transition probability can be expressed as a path integral using the Martin–Siggia–Rose formalism\footnote{More details are provided in the previous paper, \cite{lahoche2023stochastic}.} \cite{ZinnJustinBook2, livi2017nonequilibrium}:

\begin{equation}
Z[J,\bar{J},\jmath,\bar{\jmath}]= \int d\bm q d\bm{\chi}\, e^{-\Omega^2 S[\bm q,\bm \chi]+\bm J \cdot \bm q+ \bm \jmath \cdot \bm \chi}\,, \label{generatingfunctional0}
\end{equation}
where the Martin-Siggia-Rose action is given by:

\begin{align}
\nonumber \Omega^2 S[\bm q,\bm \chi]:=\int dt d\bm g\,\bigg[\Omega\bar{\chi}(\bm g,t) \chi(\bm g,t)+&i\bar{\chi}(\bm g,t)\left(\dot{\varphi}+\Omega\delta_{\bar{\varphi}} \tilde{\mathcal{H}}\right)(\bm g, t)\\
&+i\left(\dot{\bar{\varphi}}+\Omega\delta_{\varphi}\tilde{\mathcal{H}}\right)(\bm g,t)\chi(\bm g,t) \bigg]\,.\label{classicaction0}
\end{align}

The field $\bm{\chi}$, called the \textit{response field}, arises as an intermediate field, ensuring that the action is linear with respect to the Hamiltonian. In \eqref{generatingfunctional0}, $\bm{\jmath} = (\jmath, \bar{\jmath})$ is the source for the response field, and $\bm{J}(t) = {\bar{J}(\bm{g}, t), J(\bm{g}, t)}$ is the source for the complex field (conjugate to the coordinate $\bm{q}$). Moreover, the dot product is defined as:
\begin{equation}
\bm J \cdot \bm q := \int dt d\bm g \bar{\varphi}(\bm g,t) J(\bm g,t) +\int dt d\bm g \bar{J}(\bm g, t) \varphi(\bm g,t)\,.
\end{equation}
Note that this construction assumes that the path integral is defined according to the Itô prescription, in particular fixing the value of the Heaviside function at zero as $\theta(0) = 0$, see our previous work \cite{lahoche2023stochastic} for a detailed discussion. In this paper, we focus on the Abelian model $\bm{\mathrm{G}} = U(1)$, for which the irreducible representations of the group are $e^{i p \theta}$ with $p \in \mathbb{Z}$ (the standard momenta in the Fourier series transform). In Fourier space, the Hamiltonian we consider is given by:
\begin{align}
&\tilde{\mathcal{H}}[T,\bar{T}]:= \sum_{\bm p \in \mathbb{Z}^5} \int_{-\infty}^{+\infty} d\omega \, \bar{T}_{\bm p}(\omega)(\bm p^2+m^2) T_{\bm p}(\omega)\\\nonumber
&+\frac{\lambda}{2\pi} \sum_{\ell=1}^5 \sum_{\{\bm p_i \}} \int \prod_{i=1}^4 d\omega_i \delta(\omega_1+\omega_3-\omega_2-\omega_4) \mathcal{W}^{(\ell)}_{\bm p_1,\bm p_2,\bm p_3,\bm p_4} T_{\bm p_1}(\omega_1) \bar{T}_{\bm p_2}(\omega_2) T_{\bm p_3}(\omega_3) \bar{T}_{\bm p_4}(\omega_4)\,,
\end{align}
where $\bm p^2:=\sum_{\ell=1}^d p_{\ell}^2$, 
\begin{equation}
\mathcal{W}^{(\ell)}_{\bm p_1,\bm p_2,\bm p_3,\bm p_4}:=\delta_{p_{1\ell}p_{4\ell}}\delta_{p_{2\ell}p_{3\ell}} \prod_{j \neq i} \delta_{p_{1j}p_{2j}} \delta_{p_{3j}p_{4j}}\,.\label{quartickernel}
\end{equation}
and the Fourier transform is defined as:
\begin{equation}
\varphi(\bm g,t)=\int_{-\infty}^{+\infty} \frac{d\omega}{\sqrt{2\pi}} e^{-i\omega t}\sum_{\bm p \in \mathbb{Z}^d} T_{\bm p}(\omega) \prod_{\ell=1}^d e^{i p_\ell \theta_\ell}\,.
\end{equation}
The Kronecker delta involved in the definition of the \textit{quartic kernel} ensures that the quartic interaction is invariant under unitary transformations acting independently on the different group variables of the field, namely:
\begin{equation}
\varphi(\bm g) \to \varphi^\prime(\bm g):=\int d\bm{g}^\prime\, \left[\prod_{\ell=1}^d U_\ell(g_\ell,g^\prime_\ell) \right] \varphi(\bm{g}^\prime)\,,\label{unitarytrans}
\end{equation}
The matrices $U_\ell$, with entries $U_\ell(g_\ell, g^\prime_\ell)$, are assumed to be unitary, satisfying $U U^\dagger = I$. It has long been established that this property defines a regular non-locality for the interactions, known as “tensoriality,” and the terms appearing in the expansion of $\mathcal{H}$ are \textit{tensorial invariants} \cite{carrozza2015discrete}. A theorem proved in \cite{samary2014just} shows that for $d = 5$ and $\bm{\mathrm{G}} = U(1)$ (the simplest Abelian Lie group), the Euclidean quantum field theory corresponding to equilibrium states is just-renormalizable and corresponds to the so-called \textit{quartic melonic $T_5^4$} model in the literature. In particular, for melonic diagrams, which are the leading-order contributions in such a field theory, the Feynman amplitudes $A(G)$ for a melonic graph $G$ scale superficially as $A(G) \sim \Lambda^{\omega_{\text{melon}}}$ for some UV cutoff $\Lambda$, where the superficial divergence index is:
\begin{equation}
\boxed{
\omega_{\text{melon}}=4-N(G)\,,}
\end{equation}
where $N(G)$ denotes the number of external edges. Note that the melonic bound also serves as an upper bound since the number of external edges is fixed. Moreover, only the melonic graphs require renormalization. Standard reviews and papers on melonic graphs and perturbative renormalization (including proofs of BPHZ theorems) exist in this context; see, for instance, \cite{carrozza2015discrete, carrozza2014renormalization2, Lahoche:2018oeo}.

\begin{figure}
\begin{center}
\includegraphics[scale=1]{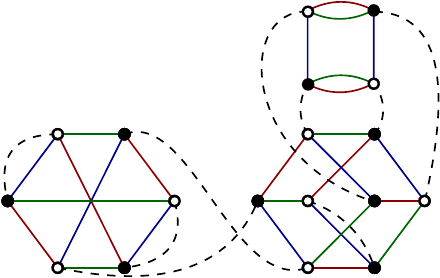}
\end{center}
\caption{A typical Feynman graph for $d=3$, involving 3 bubble vertices.}\label{figFeynman}
\end{figure}

Feynman graphs in this setting admit a convenient representation, which we summarize here. First, vertices can be represented as a $d$-colored bipartite regular graph, where black and white nodes correspond to the fields $\varphi$ and $\bar{\varphi}$, respectively, and colored edges represent Kronecker delta contractions in the definition of the interaction. Explicitly, for $d = 3$:
\begin{equation}
\vcenter{\hbox{\includegraphics[scale=0.75]{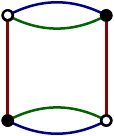}}}\equiv \int \prod_{i=1}^3 dg_i dg_i^\prime \varphi(g_1,g_2,g_3) \bar{\varphi}(g_1,g^\prime_2,g^\prime_3) \varphi(g^\prime_1,g^\prime_2,g^\prime_3)\bar{\varphi}(g_1^\prime,g_2,g_3)\,.
\end{equation}
Such an interaction pattern of a connected graph is called a \textit{bubble vertex}, or simply a bubble. Figure \ref{figFeynman} shows a typical Feynman graph for $d = 3$, where dashed edges represent Wick contractions with the bare propagator.

Within these notations, it is easy to find an explicit expression for the classical action $S[\bm q,\bm \chi]$. Defining kinetic and interaction parts as:
\begin{equation}
S=: S_{\text{kin}}+S_{\text{int}}\,,
\end{equation}
we have
\begin{align}
\nonumber S_{\text{kin}}=\sum_{\bm p \in \mathbb{Z}^5} \int_{-\infty}^{+\infty} d\hat{\omega} \bigg( \bar{\chi}_{\bm p}(\hat{\omega}) \chi_{\bm p}(\hat{\omega})&+i\bar{\chi}_{\bm p}(\hat{\omega})\left(-i\hat{\omega}+\bm p^2+m^2\right) T_{\bm p}(\hat{\omega})\\
&+i\bar{T}_{\bm p}(\hat{\omega}) \left(i\hat{\omega}+\bm p^2+m^2\right){\chi}_{\bm p}(\hat{\omega}) \bigg)\,,
\end{align}
and:
\begin{equation}
S_{\text{int}}=\frac{i\lambda}{4\pi} \sum_{\ell=1}^d\left(\, \vcenter{\hbox{\includegraphics[scale=0.8]{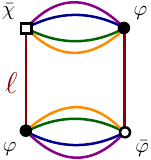}}}+\vcenter{\hbox{\includegraphics[scale=0.8]{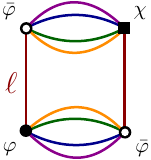}}}\,\right)\,.\label{Sint}
\end{equation}
Note that in the previous equation we introduced a graphical rule: the response fields $\chi$ and $\bar{\chi}$ are represented by black and white square nodes, respectively. Moreover, the dimensionless frequency is defined as $\omega \equiv \Omega \hat{\omega}$.\\

\medskip

To conclude, let us make a few remarks about the equilibrium bare theory. First, note that the free propagator $C$ takes the form of a $2 \times 2$ matrix, with components denoted as $C_{\chi\bar{\chi}}$, $C_{T\bar{\chi}}$, $C_{\chi\bar{T}}$, and $C_{T\bar{T}}$. The response field does not propagate, which implies that:
\begin{equation}
C_{\chi\bar{\chi}}(\hat{\omega}, \bm p^2)=0\,. \label{freepropaCHI}
\end{equation}
Other components are explicitly the following :
\begin{equation}
C_{\bar{\chi}T}(\hat{\omega}, \bm p^2)=\frac{1}{\Omega^2}\frac{\hat{\omega}-i(\bm p^2+m^2)}{\hat{\omega}^2+(\bm p^2+m^2)^2}\,,\quad C_{\bar{T}{\chi}}(\hat{\omega}, \bm p^2)=-\frac{1}{\Omega^2}\frac{\hat{\omega}+i(\bm p^2+m^2)}{\hat{\omega}^2+(\bm p^2+m^2)^2}\,,\label{freepropa1}
\end{equation}
and:
\begin{equation}
C_{T\bar{T}}(\hat{\omega}, \bm p^2)=\frac{1}{\Omega^2}\frac{1}{\hat{\omega}^2+(\bm p^2+m^2)^2}\,.\label{freepropa2}
\end{equation}

An interesting (and straightforward to prove) result is that \eqref{freepropaCHI} remains valid to all orders in perturbation theory. This is an exact statement: the intermediate (response) field does not propagate, i.e.,
\begin{equation}
\boxed{
G_{\chi \bar{\chi}}(\hat{\omega},\bm p^2)=0\,.} \label{conditionG}
\end{equation}
The proof given in \cite{aron2010symmetries2} is illuminating, and we reproduce it here. Consider the generating functional \eqref{generatingfunctional0}. We may add to the Hamiltonian $\mathcal{H}$ a linear driving term
$\frac{1}{\Omega} \int dt \sum_{\bm p} \bar{k}{\bm p}(t) T{\bm p}(t) + \text{c.c.}$
It is straightforward to see that this is equivalent to shifting the sources $\jmath$ and $\bar{\jmath}$ as:
\begin{equation}
\jmath_{\bm p} \to \jmath_{\bm p}-ik_{\bm p}\,,\quad \bar{\jmath}_{\bm p} \to \bar{\jmath}_{\bm p}+i\bar{k}_{\bm p}\,.
\end{equation}
By construction, the partition function is normalized to unity, namely $Z[J=0,\bar{J}=0]=1$. But is that case, this condition becomes $Z[0,0,-ik,i\bar{k}]=1$, and then
\begin{equation}
G_{\chi \bar{\chi}}= - \frac{\delta^2}{\delta k_{\bm p} \delta \bar{k}_{\bm p^\prime}} Z[0,0,-ik,i\bar{k}]\equiv - \frac{\delta^2 1}{\delta k_{\bm p} \delta \bar{k}_{\bm p^\prime}} =0\,.
\end{equation}

Finally, it is easy to check that:

\begin{equation}
C_{\bar{\chi}T}(\hat{\omega}, \bm p^2)-C_{\bar{\chi}T}(-\hat{\omega}, \bm p^2)=2 \hat{\omega} C_{\bar{T}T}(\hat{\omega}, \bm p^2)\,,
\end{equation}
which is the zero order version of an exact result called fluctuation dissipation theorem (FDT), coming from time reflexion symmetry and relating exact $2$-points functions components:
\begin{equation}
G_{\bar{\chi}T}(\hat{\omega}, \bm p^2)-G_{\bar{\chi}T}(-\hat{\omega}, \bm p^2)=2 \hat{\omega} G_{\bar{T}T}(\hat{\omega}, \bm p^2)\,.\label{FPTF}
\end{equation}
See reference \cite{aron2010symmetries2} for details. Recall that until this point, we choose $\Omega=1$.\\

\section{Functional renormalization group}

\subsection{General formalism}

The general objective of the functional renormalization group (FRG) is to construct a continuous interpolation between the classical action $S$ and the effective action $\Gamma$, gradually incorporating the effects of quantum fluctuations through a redefinition of the couplings and fields. Comprehensive reviews on the subject can be found in \cite{Delamotte_2012,Morris_1994a,MORRIS_1994,Berges_2002,Dupuis_2021}, while further details are provided in our previous work \cite{lahoche2023stochastic}. \\

The interpolation, generally denoted as $\Gamma_k$, corresponds physically to the effective action for the integrated out modes. The standard procedure to construct it is the following. First, we add to the classical action $S$ a regulator of the form::
\begin{equation}
\Delta S_k= \sum_{\bm p\in \mathbb{Z}^d}\sum_{a,b} \,\int_{-\infty}^{+\infty} d\omega \,\bar{\Xi}_{a}(\bm p,\omega) R_{ab,k}(\bm p,\omega) {\Xi}_{b}(\bm p,\omega)\,,
\end{equation}
where ${\Xi}(\bm p,\omega)=(\chi_{\bm p}(\omega),T_{\bm p}(\omega))$. The \textit{regulator function} $R_{ab,k}(\bm p,\omega)$ is assumed to be a differentiable function of $k$, $\bm p$ and $\omega$ ($k\in [0,\Lambda]$ for some UV cut-off $\Lambda$. Physically, it behaves as a scale-dependent mass and is designed such that high energy modes concerning the scale $k$ (i.e., such that $\hat{\omega}/k^2, \bm p^2/k^2 \ll 1$) receive a small mass whereas low energy modes decouple from IR physics. The effective average action $\Gamma_k$ is then defined as:
\begin{equation}
\Gamma_k[\bm M, \bm \sigma]+\Delta S_k[\bm M, \bm \sigma]=\bm M \cdot \bm J+ \bm \sigma \cdot \bm \jmath - W_k[\bm J, \bm \jmath\,]\,,\label{defeffectiveaction}
\end{equation}
where ${\bm M}:=(\bar M, M)$ and ${\bm \sigma}:=(\bar\sigma,\sigma)$ denote the classical fields, which are formal statistical averages:
\bea
M:=\frac{\delta W_k}{\delta  \bar J},\quad  \sigma:=\frac{\delta W_k}{\delta  \bar j}.
\eea

The fundamental equation of the FRG is the so-called Wetterch-Morris equation \cite{Delamotte_2012}:
\begin{equation}
\frac{\partial}{\partial k} \Gamma_k= \Tr\, \frac{\partial \bm R_k}{\partial k} (\bm \Gamma_k^{(2)}+\bm R_k)^{-1}\,,\label{Wett}
\end{equation}
where capital bold letters denote $2\times 2$ matrix-valued functions and the trace runs over all field indices. This equation describes how the effective average action evolves as the IR cutoff $k$ decreases from the UV to the IR, such that $\Gamma_\Lambda \simeq S$ (the classical action, with no fluctuations integrated out) and $\Gamma_{k=0}\equiv \Gamma$ (the full effective action, with all fluctuations integrated out). Note that, because of the regulator, it is appropriate to consider the continuum limit $\Lambda \to \infty$ in the flow equation \eqref{Wett}. \\

\subsection{Regulator and truncation}
In order to preserve the condition that the response field does not propagate along the RG flow, we assume the regulator to be of the following form in matrix notation,
\begin{equation}
\bm R_k(\bm p,\omega):= \begin{pmatrix}
R_{k}^{(2)}(\bm p,\omega) & +i R_k^{(1)}(\bm p,\omega)\\
i R_k^{(1)}(\bm p,-\omega) &0
\end{pmatrix}\,, \label{equationregul}
\end{equation}
where Fourier components are defined such as:
\begin{equation}
R_k(\bm p,t):=\frac{1}{2\pi} \int d\omega\, e^{-i\omega t} R_k(\bm p,\omega)\,.
\end{equation}
Time reversal symmetry imposes a constraint on the definition of the regulator function, derived in our previous work \cite{lahoche2023stochastic} but inspiring from \cite{duclut2017frequency}:
\begin{equation}
R_k^{(1)}(\bm p,t^\prime-t)-R_k^{(1)}(\bm p,t-t^\prime)-\frac{2}{\Omega}\dot{R}_{k}^{(2)}(\bm p,t^\prime-t)=0\,. \label{relationregulator}
\end{equation}
To simplify the presentation of the general methodology, we assume that no coarse-graining is applied to the frequency spectrum and that the regulator function depends only on $\bm{p}^2$. Under this assumption, the previous equation shows that we can set ${R}_{k}^{(2)} = 0$. In this way, causality and time-reversal symmetry are not explicitly broken by the regulator.\\

The exact flow equation \eqref{Wett} cannot be solved exactly and require approximations. In our recent investigation \cite{lahoche2023stochastic}, we considered the following approximation:

\begin{align}
\nonumber \Gamma_k[M,\bar{M},\sigma,\bar{\sigma}]&=\sum_{\bm p \in \mathbb{Z}^5} \int_{-\infty}^{+\infty} d\hat{\omega} \bigg(Y(k) \bar{\sigma}_{\bm p}(\hat{\omega}) \sigma_{\bm p}(\hat{\omega})\\\nonumber
&+i\bar{\sigma}_{\bm p}(\hat{\omega})\left(-iY(k)\hat{\omega}+Z(k)\bm p^2+m^2(k)\right)M_{\bm p}(\hat{\omega})\\\nonumber
&+i \bar{M}_{\bm p}(\hat{\omega}) \left(iY(k)\hat{\omega}+Z(k)\bm p^2+m^2(k)\right){\sigma}_{\bm p}(\hat{\omega})\\
&+i \bigg(\bar{\sigma}_{\bm p}(\hat{\omega}) \frac{\delta \hat{\mathcal{H}}_{\text{int},k}}{\delta \bar{M}_{\bm p}(\hat{\omega})}+{\sigma}_{\bm p}(\hat{\omega}) \frac{\delta \hat{\mathcal{H}}_{\text{int},k}}{\delta {M}_{\bm p}(\hat{\omega})} \bigg)\bigg)\,,\label{AnsatzGamma}
\end{align}
where the kinetic Hamiltonian $\mathcal{H}_{\text{int}}$ looks as a local expansion:
\begin{equation}
\hat{\mathcal{H}}_{\text{int},k}[M,\bar{M}]=\int d\hat{t}\sum_b k^{\dim(b)} Z^{n(b)}(k)\bar{\lambda}_b \Tr_b[\varphi(\hat{t}),\bar{\varphi}(\hat{t})]\,,\label{defexpansion}
\end{equation}
the sum runs over bubbles $b$, with the symbol $\Tr$ denoting the unitary-invariant contraction of the field indices. Moreover, $\dim(b)$ is the scaling dimension of the bubble $b$, computed from the equilibrium theory. One can show that, even in the context of a field theory without a background space-time, the dimension must be defined as \cite{Lahoche:2018oeo, Carrozza_2017a,carrozza2015discrete}:
\begin{definition}
Let $b$ a bubble having $n(b)$ white vertices and $\mathbb{G}$ the set of $2$-points diagrams made of a single vertex of type $b$. The scaling dimension $\dim(b)$ is defined as:
\begin{equation}
\dim(b)=2-\underset{r\in \mathbb{G}}{\max} \,\,\omega(r)\,.\label{canonicaldim}
\end{equation}
\end{definition}
In particular, it is straightforward to verify that, for the rank-five Abelian model, the mass has dimension $2$, while the melonic quartic bubble is dimensionless, with all other interactions being irrelevant. In \eqref{AnsatzGamma}, $Y(k)$, $Z(k)$, and $m^2(k)$ denote the wave-function renormalization and the effective mass, respectively. The chosen parametrization preserves time-reversal symmetry, corresponding to the transformation:
\begin{equation}
T_{\bm p}^\prime(t)=T_{\bm p}(-t)\,,\quad \chi^\prime_{\bm p}(t)=\chi_{\bm p}(-t)+\frac{2i}{\Omega} \dot{T}_{\bm p}(-t)\,. \label{transT1}
\end{equation}

Before coming on the computations, let us precise some points. First of all, we will assume the regulator to be of the form:
\begin{equation}
R^{(1)}_k(\bm p)= k^2Z(k) r_k(\bm p^2) \,,
\end{equation}
where for $r_k(\bm p^2)$ we choose the standard Litim regulator \cite{litim2000optimisation}:
\begin{equation}
r_k(\bm p^2):= \left(1-\frac{\bm p^2}{k^2}\right)\theta(k^2-\bm p^2)\,.
\end{equation}

\begin{remark}
The definition of the regulator above imposes a bound on the anomalous dimension. Indeed, suppose that $\eta_*$ is the value reached at some fixed point, we have $Z(k)=k^{\eta_*}$ by definition of the anomalous dimension. The $k$ dependency of the regulator then became $k^{2+\eta_*}$, and the condition $\lim_{k\to \infty} R^{(1)}_k(\bm p) \to \infty$, imposes:
\begin{equation}
\eta_* >-2\,.\label{boundeta}
\end{equation}
\end{remark}
\subsection{The non-branching melons theory space}

The tensorial theory space is highly intricate, but several subregions stable under renormalization have been identified by various authors. In particular, the non-branching melonic sector is stable in the deep UV regime, defined as:
\begin{equation}
\Lambda \ll k \ll 1,
\end{equation}
has been considered in \cite{Carrozza_2017,Carrozza_2017a,pithis2020phase}, and in \cite{Lahoche:2018oeo}, as a sector where the infinite hierarchy of the flow equation can be closed, using the non trivial Ward identities (see the discussion below). Here, we recall the definition of the cyclic melonic sector. \\

For $d$-colored graphs, melonic graphs can be defined recursively \cite{Gurau_2016}. The basic piece is the elementary melon, namely:

\begin{equation}
b_1:=\vcenter{\hbox{\includegraphics[scale=1]{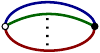} }}\,.
\end{equation}
From $b_1$, any melon can be defined as follow:
\begin{definition}\label{defmelons}
Any melonic bubble $b_\kappa$ of valence $\kappa$ may be deduced from the elementary melon $b_1$ by replacing successively $\kappa-1$ colored edges (including maybe color ‘‘0") by $(d-1)$-dipole, the $(d-1)$-dipole insertion operator $\mathfrak{R}_{i}$ being defined as:
\begin{equation}
\vcenter{\hbox{\includegraphics[scale=1]{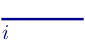} }}\underset{\mathfrak{R}_{i}}{\longrightarrow}\vcenter{\hbox{\includegraphics[scale=1]{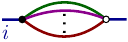} }}\,.
\end{equation}
In formula: $b_\kappa := \left(\prod_{\alpha=1}^{\kappa-1}\mathfrak{R}_{i_\alpha}\right) b_1$.
\end{definition}
Non-branching melons are now defined as:
\begin{definition}\label{defnonbranch}
A non-branching melonic bubble of valence $\kappa$, $b_\kappa^{(\ell)}$ is labeled with a single index $\ell\in\llbracket 1,5\rrbracket$, and defined such that:
\begin{equation}
b_\kappa^{(\ell)}:= \left(\mathfrak{R}_{\ell}\right)^{\kappa-1}\,b_1\,.
\end{equation}
\end{definition}
Figure \ref{fig2} provides the generic structure of melonic non-branching bubbles in rank $3$. Note that the definition holds for diagrams involving square nodes.
\begin{center}
\begin{equation*}
\vcenter{\hbox{\includegraphics[scale=1]{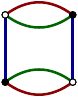} }} \,\underset{\mathfrak{R}_{i}}{\longrightarrow}\, \vcenter{\hbox{\includegraphics[scale=1]{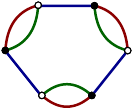} }}\,\cdots \underset{\mathfrak{R}_{i}}{\longrightarrow}\, \vcenter{\hbox{\includegraphics[scale=1]{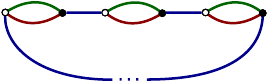} }}\underset{\mathfrak{R}_{i}}{\longrightarrow}\cdots
\end{equation*}
\captionof{figure}{Structure of the non-branching melons, from the smallest one $b_2$.} \label{fig2}
\end{center}

\subsection{Boundary of a Feynman graph}

The last ingredient needed to construct the $\beta$-function is the concept of the boundary graph, which we briefly review in this section. This concept is particularly useful for identifying the contributions of the different bubbles that appear on both sides of the flow equation (see below). The definition is straightforward and is illustrated in Figure \ref{figFeynman2}:

\begin{definition}\label{defboundary}
Let $G$ be a regular $(d+1)$-colored Feynman diagram with $2N$ external dotted edges. These edges are attached to $2N$ black and white nodes, called external nodes. The boundary diagram $\partial G$ of $G$ is then defined as the regular $d$-colored graph obtained by discarding the edges of color $0$ and such that:
\begin{enumerate}
\item Nodes of $\partial G$ are external nodes of $G$
\item Edges with color $\neq 0$ linking two external nodes are conserved.
\item Any open cycle made of colors $0$ and $i$ between two external nodes $n$ and $\bar{n}$ is replaced by a link of color $i$ in $\partial G$.
\end{enumerate}
\end{definition}
Note that the boundary diagram is melonic, but not in the non-branching sector in the example provided by the Figure. 

\begin{center}
\begin{equation*}
\vcenter{\hbox{\includegraphics[scale=0.8]{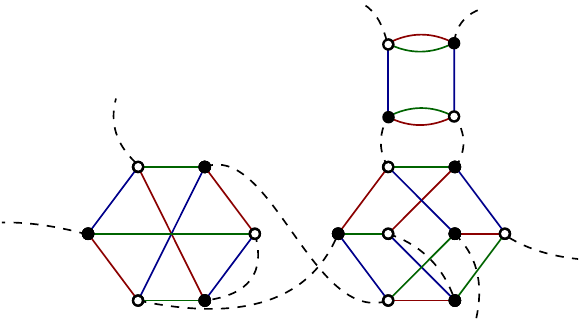}}}\quad \Huge{\rightarrow} \quad \vcenter{\hbox{\includegraphics[scale=1]{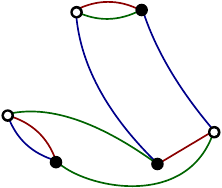}}}
\end{equation*}
\captionof{figure}{Illustration of the mapping $G\to \partial G$ for a 6-points Feynman diagram in rank 3.}\label{figFeynman2}
\end{center}

\subsection{Method for computing $\beta$-functions}\label{method}

Before closing this section, let us illustrate how to compute the $\beta$-functions. In this paper, we focus on the symmetric phase, that is, we expand the effective average action in powers of the classical field. Moreover, the truncation is taken to be linear in the response field, which is expected to be a good approximation in the equilibrium dynamics phase (see below and \cite{canet2011general}). In general, tensorial methods are restricted to expansions around the symmetric phase (see \cite{Benedetti_2015,Benedetti_2016,Ben_Geloun_2015}), although expansions beyond the symmetric phase have also been considered in certain regimes (see, for instance, \cite{pithis2021no,pithis2020phase}). Here, we illustrate the general strategy for the truncation described above, noting that the methodology is the same in general.\\

The first step is to take successive functional derivative with respect to classical fields $M, \bar{M}, \sigma, \bar{\sigma}$. We introduce the compact notation $\Xi=\{M,\sigma \}$, $\bar{\Xi}=\{\bar{M},\bar{\sigma}\}$ and we define:
\begin{equation}
\Gamma_{k,\bar{\Xi}^{a_1}\cdots \bar{\Xi}^{a_P}\cdots {\Xi}^{b_1} \Xi^{b_P}}^{(2P)}=\frac{\delta^{2P} \Gamma_k}{\delta \bar{\Xi}^{a_1}_{\bm p_1}(\hat{\omega}_1)\cdots \delta \bar{\Xi}^{a_P}_{\bm p_P}(\hat{\omega}_P)\cdots \delta {\Xi}^{b_P}_{\bm p_P^\prime}(\hat{\omega}_P^\prime)}\,,
\end{equation}
for $a_i,b_i=0,1$, $\Xi^0=M$, $\Xi^1=\sigma$. The truncation \eqref{AnsatzGamma} gives for instance:
\begin{equation}
\Gamma_{k, \bar{\sigma} \sigma}^{(2)}=Y(k) \delta_{\bm p_1\bm p_2} \delta(\hat{\omega}_1-\hat{\omega}_2)\,,
\end{equation}
and:
\begin{equation}
\Gamma_{k, \bar{\sigma} M \bar{M} M}^{(4),(\ell)}= \frac{i}{\pi} \pi^{(2)}_k(p_{1\ell}^2,p_{3\ell}^2)\left(\mathcal{W}^{(\ell)}_{\bm p_1,\bm p_2,\bm p_3,\bm p_4} +\bm p_2 \leftrightarrow \bm p_4\right)\delta(\hat{\omega}_1-\hat{\omega}_2+\hat{\omega}_3-\hat{\omega}_4)\,.\label{rencondGamma4}
\end{equation}
The vertex function $\pi^{(2)}_k(p_{1\ell}^2,p_{3\ell}^2)$, including in principle the full dependency on the vertex on external momenta, is normalized such that:
\begin{equation}
\pi^{(2)}_k(0,0)=: \lambda(k)\,, \label{rencondcoupling}
\end{equation}
is the effective quartic coupling at scale $k$. Note that, in the symmetric phase, the flow equation for $Y$ vanishes exactly, but not the one of $Z$ (what will be clear below, but more details can be found in \cite{lahoche2023stochastic}). Because the $2$-point function play a important role in the computation, it is suitable to introduce the notations:
\begin{equation}
\Gamma_{k,\bar{\Xi}^{a_1}\Xi^{a_2}}^{(2)}=: \gamma_{k,\bar{\Xi}^{a_1}\Xi^{a_2}}^{(2)}(\bm{p}_1,\hat{\omega}_1) \delta_{\bm{p}_1\bm{p}_2}\delta(\hat{\omega}_1-\hat{\omega}_2)\,,\label{decomp2points}
\end{equation}
and we have, because of the definition \eqref{AnsatzGamma}:
\begin{equation}
\gamma_{k,\bar{\sigma}M}^{(2)}(\bm{p}=\bm{0},\hat{\omega}_1=0)=i\, m^2(k)\,,\label{defmass}
\end{equation}
and:
\begin{equation}
\frac{d}{d\hat{\omega}_1}\gamma_{k,\bar{\sigma}M}^{(2)}(\bm{p}=\bm{0},\hat{\omega}_1=0):=Y(k)\,,\qquad \frac{d}{dp^2_i}\gamma_{k,\bar{\sigma}M}^{(2)}(\bm{p}=\bm{0},\hat{\omega}_1=0):=iZ(k)\,.\label{defZY}
\end{equation}

\begin{figure}
\begin{center}
\includegraphics[scale=1]{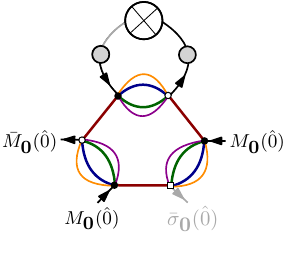}\qquad \includegraphics[scale=1]{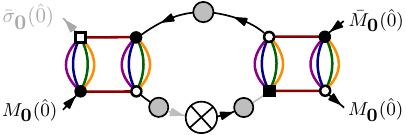}
\end{center}
\caption{Typical contributions involved on the right hand side of the flow equation for $\lambda$.}\label{figRHSbeta4}
\end{figure}

Now, let us illustrate how that work with some details. Taking the derivatives with respect to $\bar{\sigma}$ and $M$ on both sides of the flow equation, we get, setting all the fields to zero on both sides:
\begin{equation}
\dot{\gamma}_{k,\bar{\sigma}M}^{(2)}(\bm{p}_1,\hat{\omega}_1)\delta_{\bm{p}_1\bm{p}_2}\delta(\hat{\omega}_1-\hat{\omega}_2)=-\Tr\,\dot{\bm R}_{k} \bm G_{k} \Gamma_{k\bar{\sigma} M\bullet \bullet }^{(4)} \bm G_{k}\,,\label{eqflowMass}
\end{equation}
the trace $\Tr$ running over momenta, frequencies and fields, and the dots in the $4$-point function is for the field involved in the trace. Moreover, the notation "dot" is for the derivation with respect to $\ln(k/\Lambda)$. The propagator $G$ is a $2\times 2$ matrix, that reads 
\begin{equation}
\bm G_{k}:=(\bm \Gamma^{(2)}+\bm R_k)^{-1}\,,
\end{equation}
and in the symmetric phase, the components are explicitly:
\begin{equation}
G_{k,\bar{\sigma} M}(\bm p^2,\hat{\omega})=- \frac{1}{k^2}\frac{i}{Z(k)}\frac{1}{\hat{f}(x,-y)}\label{propa1}
\end{equation}
and:
\begin{equation}
G_{k,\bar{M} M}(\bm p^2,\hat{\omega})=\frac{1}{k^4}\frac{1}{Z^2(k)}\frac{1}{\hat{f}(x,y)\hat{f}(x,-y)}\,,\label{propa2}
\end{equation}
where, in the derivative expansion approximation:
\begin{equation}
\hat{f}(x,y)=iy+x+\bar{m}^2+ r(x)\,, \label{truncationf}
\end{equation}
and:
\begin{equation}
\bm p^2 = k^2 x\,,\qquad \omega =  Z(k) k^2 Y^{-1}(k) y\,.
\end{equation}
Finally, barred quantities (dimensionless), are defined such that:
\begin{equation}
m^2=: Z(k) k^2 \bar{m}^2\,,\qquad {\pi}_k^{(2)}= Z^2(k) \bar{\pi}_k^{(2)}\,,
\end{equation}
in agreement with the power counting. Graphically, the resulting flow equation for the 2-point function reads:
\begin{align}
\nonumber\dot{\gamma}_{k,\bar{\sigma}M}^{(2)}(\bm{p}_1,\hat{\omega}_1)\delta_{\bm{p}_1\bm{p}_2}\delta(\hat{\omega}_1-\hat{\omega}_2)&=-\sum_{i=1}^d\Bigg(\vcenter{\hbox{\includegraphics[scale=1]{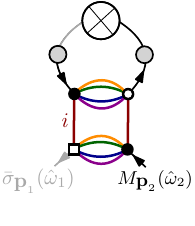}}}+\vcenter{\hbox{\includegraphics[scale=1]{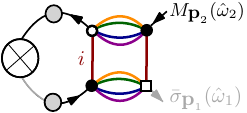}}}\Bigg)\,,\label{flowmassdiag}
\end{align}
the conventions for the edges being the following:
\begin{equation}
G_{\bar{M}M}:=\vcenter{\hbox{\includegraphics[scale=1.2]{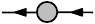}}}\,,\qquad G_{\bar{\sigma}M}:=\vcenter{\hbox{\includegraphics[scale=1.2]{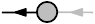}}}\,.
\end{equation}
In \eqref{flowmassdiag}, only the two first ones diagrams on the right hand side are melonic, and in the deep UV regime we consider, they dominate the flow, and we discard the non melonic contributions. The diagrams can be computed explicitly, for instance:
\begin{align}
\nonumber \vcenter{\hbox{\includegraphics[scale=0.8]{OneLoopMass2.pdf}}}&=i \bar{\pi}_k^{(2)}(p_{11}^2,p_{11}^2) \delta_{\bm{p}_1\bm{p}_2}\delta(\hat{\omega}_1-\hat{\omega}_2) Z(k) k^2 L_{21}(x_1)\,,
\end{align}
where $x_1:= p_1^2/k^2$ and assuming $k$ large enough sums can in the definition of $L_{21}(x_1)$ be replaced by integrals. Setting external momenta in \eqref{flowmassdiag} to zero, we get because of the definition \eqref{defmass}:
\begin{equation}
\dot{\bar{m}}^2=-(2+\eta){m}^2-d\bar{\lambda}\,L_{21}(0)\,,\label{betam0}
\end{equation}
where:
\begin{equation}
L_{21}(x_1):= \frac{4}{\pi} \int_{\mathbb{R}^{d+1}} d\textbf{x}^\prime dy \delta(x_1^\prime-x_1) \mu_1(x^\prime,y) \frac{1+\hat{\tau}(y)r(x^\prime)}{\hat{f}(x^\prime,y)\hat{f}^2(x^\prime,-y)}\,,
\end{equation}
and the anomalous dimension is:
\begin{equation}
\eta:= \frac{1}{Z(k)} k \frac{d}{dk} Z(k)\,.
\end{equation}
The equation for the quartic coupling $\lambda$ can be derived in the same way. In this case, however, the right-hand side involves diagrams such as those shown in Figure \ref{figRHSbeta4}. Particular attention must be paid to the edge diagrams: the rule is that only those diagrams whose edge diagram corresponds to the interaction whose flow is being studied should be retained on the right-hand side.\\

The diagrams involve sextic couplings (at zero momenta):
\begin{equation}
\Gamma_{k, \bar{\sigma} M \bar{M}M \bar{M} M}^{(6),(\ell)}\bigg\vert_{\text{0}}=\frac{9i}{\pi^2}\kappa \delta(\hat{\omega}_1-\hat{\omega}_2+\hat{\omega}_3-\hat{\omega}_4+\hat{\omega}_5-\hat{\omega}_6)\,,\label{gamma6zero}
\end{equation}
or are quadratic with respect to the quartic coupling. Each diagram can be computed in the continuum limit, and we get:
\begin{align}
\beta_\lambda=-2\eta \bar{\lambda}-\frac{3}{2}\bar{\kappa}\left(L_{21}(0)-L_{22}(0) \right)+\frac{8\bar{\lambda}^2}{\pi} \left(L_{31}+\frac{1}{2}L_{32} \right)\,,
\end{align}\label{flowphi4}
where:
\begin{align}
L_{32}&:= \int_{\mathbb{R}^{4+1}} d\textbf{x} dy \mu_1(x,y)\frac{1}{f^2(x,y)f^2(x,-y)}\,,\\
L_{31}&:= \int_{\mathbb{R}^{4+1}} d\textbf{x} dy \frac{\mu_1(x,y)}{f^2(x,y)}\frac{1}{f(x,y)f(x,-y)}\,.
\end{align}

The last piece is the anomalous dimension, and from the definition \eqref{defZY}:
\begin{equation}
\eta=-\bar{\lambda}^{\prime}L_{21}(0)-\bar{\lambda} \frac{d}{d p_1^2} L_{21}(x_1)\big\vert_{x_1=0}\,,\label{eqeta}
\end{equation}
where the first term is the derivative of the effective vertex:
\begin{equation}
\frac{d}{dp_1^2} \pi_k^{(2)}(p_1^2,p_1^2)\bigg\vert_{p_1=0}=:Z^2(k)k^{-2}\bar{\lambda}^{\prime}\,.\label{derivvertex}
\end{equation}
A distinctive feature of tensorial field theories is that $\eta$ exhibits a nontrivial flow in the symmetric phase, which is exceptional and arises from the non-locality of the interactions (the flow of $Y$ vanishes because the interactions are local in time). \\

In principle, this procedure can be continued indefinitely, with the flow of the sextic coupling generating an octic coupling, and so on. The method we presented in \cite{lahoche2023stochastic}, as a continuation of \cite{Lahoche:2018oeo,Lahoche_2020b}, allows this infinite hierarchy to be closed, and the otherwise elusive function $\lambda^\prime$ can be computed. The main steps of this construction are recalled in the next section, and they form the basis for the remainder of this paper. In summary:

\begin{enumerate}
    \item The effective vertex function $\kappa$ can be computed using the Schwinger-Dyson equations in the melonic sector.
    
    \item The function $\lambda^\prime$ can be fixed non-perturbatively using the non-trivial Ward identities arising because of the explicit symmetry breaking of the bare propagator.
\end{enumerate}

\section{Highlights of stochastic dynamics in group field theory}\label{summary}

In this section, we review the results obtained in the equilibrium dynamics regime in our previous work \cite{lahoche2023stochastic}, both because these results are useful for later developments and to make this article sufficiently self-contained to be read independently of the first. In particular, we first focus on establishing the explicit expression for $\kappa$. Since the initial theory is quartic, the sextic interactions are purely effective and can be constructed, for instance, perturbatively. The recursive structure of the melonic diagrams imposes strong constraints on the structure of the Feynman graphs, which resemble trees and can be easily resummed. Equivalently, the same results can be obtained from the Schwinger-Dyson equations (see \cite{lahoche2021no,Lahoche:2018oeo,Lahoche_2020b,samary2014closed} and the appendix of \cite{lahoche2023stochastic}).\\

We consider on the set of Feynman graphs $S=\{ \mathcal{F}^{-1}[\bar{G}_0] \vert \partial G_i=\partial G_0 \,\forall G_i\in \mathcal{F}^{-1}[\bar{G}_0]\}$, where the boundary $\partial G_0$ corresponds to the cyclic sectic melonic interaction:
\begin{equation}
\partial G_0:=\vcenter{\hbox{\includegraphics[scale=1]{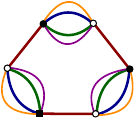}}}\,.
\end{equation}
Hence, the zero momenta effective vertex function $\Gamma_{k, \bar{\sigma} M \bar{M}M \bar{M} M}^{(6),(\ell)}\big\vert_{\text{0}}$ must decomposes as:
\begin{equation}
\Gamma_{k, {\sigma}\bar{M}M \bar{M} M\bar{M}}^{(6),(\ell)}\big\vert_{\text{0}}= 12\times \sum_{G\in S} \mathcal{A}_G\,,
\end{equation}
where $\mathcal{A}_G$ denotes Feynman amplitude for the graph $G$. It is easy to check that the effective 6-point function must have the following structure:
\begin{equation}
\Gamma_{k, {\sigma}\bar{M}M \bar{M} M\bar{M}}^{(6),(\ell)}\big\vert_{\text{0}}= 12\times \left(\vcenter{\hbox{\includegraphics[scale=0.31]{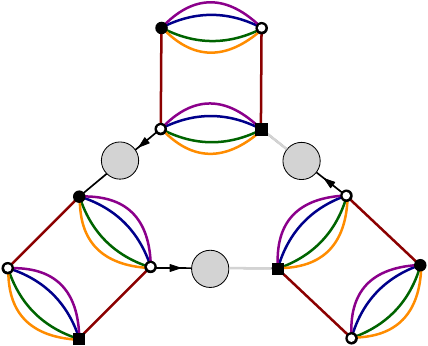}}}\quad +\quad \vcenter{\hbox{\includegraphics[scale=0.31]{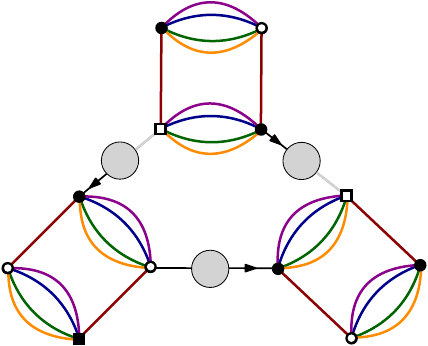}}} \quad +\quad \vcenter{\hbox{\includegraphics[scale=0.31]{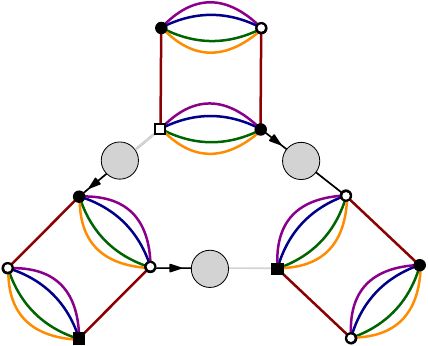}}} \right)\,,
\end{equation}
As before, gray discs represent effective propagators, while here the bubbles correspond to effective vertices. The diagrams can be easily computed using the observation in \cite{Lahoche:2018oeo} that the effective propagator can be replaced by the truncation \eqref{propa1} and \eqref{propa2} for superficially UV-convergent functions (an approximation consistent with the Ward identities). Using definition \eqref{gamma6zero}, we then obtain:
\begin{equation}
\bar{\kappa}=\frac{32\bar{\lambda}^3}{\pi} \int_{\mathbb{R}^4} d\textbf{x} \int dy\frac{1}{f^3(x,y)f(x,-y)}\,.\label{equationkappa}
\end{equation}
Explicitly:
\begin{align}
\bar{\kappa}= \frac{4\pi ^2\bar{\lambda}^3 (\bar{m}^2 (\bar{m}^2+3)+3)}{(1+\bar{m}^2)^3} \,.
\end{align}

\begin{remark}
The relevance of this approach lies in the reliability of the approximation, which consists of replacing the (superficially convergent) inner loop by its truncation-based approximation. The validity of this approximation has been discussed in \cite{Lahoche_2020b}, where an alternative method was presented that fixes the sextic interactions directly using Ward identities, a method we will employ in the following. We will return to the question of the reliability of this approximation in section \ref{numan}.
\end{remark}

The computation of $\lambda^\prime$ requires evaluating the non-trivial Ward identities that arise from the kinetic action, which breaks the unitary symmetry,
\begin{equation}
T_{\bm p} \to T_{\bm p}^\prime = \sum_{q_i\in \mathbb{Z}} U_{p_i q_i} T_{\bm q}\big\vert_{q_j=p_j\,\forall j\neq i}\,,
\end{equation}
for some unitary transformation $U$. Due to the path integration, the partition function is formally expected to be invariant. Expanding this invariance condition for an infinitesimal unitary transformation, we obtain:
\begin{equation}
\hat{\epsilon}_i[T]_{\bm p} := \sum_{q_i} \epsilon_{p_i q_i} T_{\bm q} \big\vert_{q_j=p_j\,,j\neq i}\,,
\end{equation}
where $ \epsilon_{p_i q_i}$ is a natural representation of the Lie algebra, we obtain the condition:

\begin{equation}
0\equiv \int d\bm q d\bm \chi \left[\hat{\epsilon}_i[S[\bm q,\bm \chi]]+\hat{\epsilon}_i[\Delta S_k[\bm q,\bm \chi]]-\hat{\epsilon}_i[\bm J\cdot \bm q+\bm \jmath \cdot \bm \chi] \right]e^{-S[\bm q,\bm \chi]-\Delta S_k[\bm q,\bm \chi]+\bm J \cdot \bm q+ \bm \jmath \cdot \bm \chi}\,.\label{var1}
\end{equation}
After some efforts, and introducing the notations:
\begin{equation}
G^{(n+\bar{n};M+\bar{M})}_{p_1\cdots p_n, \bar{p}_1\cdots \bar{p}_{\bar{n}}; p_I\cdots p_M, \bar{p}_1\cdots \bar{p}_{\bar{M}}}:=\prod_{i=1}^n \frac{\partial}{\partial \bar{J}_{p_i}} \prod_{\bar{i}=1}^{\bar{n}} \frac{\partial}{\partial {J}_{\bar{p}_{\bar{i}}}} \prod_{I=1}^M \frac{\partial}{\partial \bar{\jmath}_{p_I}} \prod_{\bar{I}=1}^{\bar{M}} \frac{\partial}{\partial {\jmath}_{\bar{p}_{\bar{I}}}}W_k\,,
\end{equation}
and:
\begin{equation}
M_{p}:=\frac{\partial W_k}{\partial \bar{J}_p} \,,\quad \bar{M}_{p}:=\frac{\partial W_k}{\partial {J}_p}\,, \quad \sigma_{p}:=\frac{\partial W_k}{\partial \bar{\jmath}_p}\,,\quad \bar{\sigma}_{p}:=\frac{\partial W_k}{\partial {\jmath}_p}\,,
\end{equation}
one can obtain the following statement:
\begin{proposition}\label{propositionWard}
Observable of the equilibrium dynamical model satisfy the following Ward-Takahashi identity:
\begin{align}
\nonumber0&=\int d{\omega}\sum_{\bm p, \bm p^\prime} \prod_{j\neq i} \delta_{p_jp_j^\prime} \bigg[\left(iZ_{\infty}\left[p_i^2-p_i^{\prime 2}\right]+i[R_k^{(1)}(\bm p,\omega)-R_k^{(1)}(\bm p^\prime,\omega)]\right) \\\nonumber
&\times\bigg(G_{k,\bar{\sigma}M}^{(1;\bar{1})}(\bm p^\prime,{\omega};\bm p,{\omega})
+\bar{\sigma}_{\bm p}({\omega})M_{\bm p^\prime}({\omega})\bigg)\\\nonumber
&+\left(iZ_{\infty}\left[p_i^2-p_i^{\prime 2}\right]+i[R_k^{(1)}(\bm p,-\omega)-R_k^{(1)}(\bm p^\prime,-\omega)]\right)\\\nonumber
&\times\bigg(G_{k,\bar{M}{\sigma} }^{(\bar{1};1)}( \bm p,{\omega};\bm p^\prime,{\omega})+{\sigma}_{\bm p^\prime}({\omega})\bar{M}_{\bm p}({\omega})\bigg)\\\nonumber
&+[R_k^{(2)}(\bm p,\omega)-R_k^{(2)}(\bm p^\prime,\omega)] \left(G_{k,\bar{\sigma}\sigma}^{(0;{1}+\bar{1})}(\bm p^\prime,{\omega};\bm p,{\omega})+{\sigma}_{\bm p^\prime}({\omega})\bar{\sigma}_{\bm p}({\omega}) \right)\\
&-\bar{J}_{\bm p}({\omega}) M_{\bm p^\prime}({\omega})+ J_{\bm p^\prime}({\omega})\bar{M}_{\bm p}({\omega})-\bar{\jmath}_{\bm p}({\omega})\sigma_{\bm p^\prime}({\omega})+\jmath_{\bm p^\prime}({\omega})\bar{\sigma}_{\bm p}({\omega})
\bigg]\delta_{p_ip}\delta_{p_i^\prime p^\prime} \,.\label{Ward1}
\end{align}
\end{proposition}

Differentiating the above expression with respect to the classical fields and then setting them to zero yields relations between effective vertices and their momentum dependence. In particular, differentiating the Ward identity \eqref{Ward1} with respect to $M_{\bm q}({\omega}1)$ and $\bar{\sigma}_{\bm{\bar{q}}}({\bar{\omega}}1)$ provides a relation between $\Gamma_{k,\sigma\bar{M} M \bar{M}}^{(4)}$ and $\Gamma_{k,\bar{\sigma} M}^{(2)}$. This relation becomes more transparent when expressed in the graphical notation, which explicitly reveals the non-local structure of the interactions:
\begin{align}
\nonumber &\Bigg(\vcenter{\hbox{\includegraphics[scale=0.8]{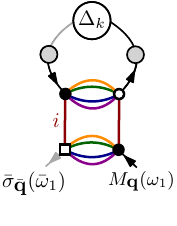}}}+\vcenter{\hbox{\includegraphics[scale=0.8]{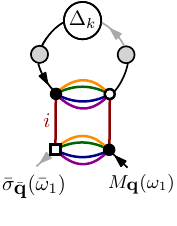}}}\Bigg)\delta_{p^\prime q_i} \delta_{p\bar{q}_i}\prod_{j\neq i}\delta_{q_j{\bar{q}}_j}\delta(\omega_1-\bar{\omega}_1)\\\nonumber
&-\sum_{\bm p, \bm p^\prime} \prod_{j\neq i} \delta_{p_jp_j^\prime} \Delta_k(\bm p,\omega_1) \delta_{\bm p\bm{\bar{q}}} \delta_{\bm q\bm p^\prime}\delta(\omega_1-\bar{\omega}_1) +\sum_{\bm p, \bm p^\prime} \prod_{j\neq i} \delta_{p_jp_j^\prime}\bigg[\gamma_{k,\bar{\sigma}M}^{(2)}(\bm p,\omega_1)- \gamma_{k,\bar{\sigma}M}^{(2)}(\bm p^\prime,\omega_1)
\bigg] \\
&\times \delta_{\bm p^\prime \bm q}\delta_{\bm p\bm{\bar{q}}} \delta_{p_ip}\delta_{p_i^\prime p^\prime}\delta(\omega_1-\bar{\omega}_1)+ \sum_{\bm p, \bm p^\prime} \prod_{j\neq i} \delta_{p_jp_j^\prime}\delta_{\bm p\bm{\bar{q}}}\delta_{\bm q\bm p^\prime} i\delta R_k^{(1)}(\bm p,\omega_1)\delta(\omega_1-\bar{\omega}_1)=0\,,
\end{align}
where:
\begin{align}
\nonumber \Delta_k(\bm{p},\omega)&\equiv
\begin{pmatrix}
0&iZ_{\infty}\delta p^2+i\delta R_k^{(1)}(\bm p,\omega)\\
iZ_{\infty}\delta p^2+i\delta R_k^{(1)}(\bm p,-\omega)&0
\end{pmatrix} \delta_{p_ip}\delta_{p_i^\prime p^\prime}\\
&=:\Delta_k^\prime(\bm{p},\omega)\delta p^2 \,,\label{deltadef}
\end{align}
the variation $\delta p^2:=p_i^2-(p_i^\prime)^2$ being at first order in the difference. Moreover, $Z_\infty$ is the field renormalization required to make the continuum limit well defined ($\Lambda\to \infty$), and is fixed by the renormalization of the equilibrium theory (see \cite{Lahoche:2018oeo}), $Z_\infty^{-1}\sim \ln \Lambda$. We set $\bm q_\bot=\bm q_\bot^\prime$, $p^\prime=q_i$, $p=\bar{q}_i$, and $p=p^\prime+1$, and make use of the continuum limit to replace finite difference by derivatives. The final result looks as a constraint in the continuum limit:

\begin{align}
\nonumber \frac{4i\bar{\lambda}(k)}{\pi} \int dy & \int_{\mathbb{R}^4} d\bm x \frac{\theta(1-x)-Z_{\infty}{Z}^{-1}(k)}{\hat{f}(x,y)\hat{f}^2(x,-y)}\to 1\,,\label{Wardcontinuum}
\end{align}
This relation has been investigated in \cite{Lahoche:2018oeo,Lahoche_2020b} and serves as an intermediate result for what follows. Note that the previous equation can give a misleading impression. In reality, the integral proportional to $Z_\infty$ cannot be accurately computed using the truncation. The same applies to the other integral, where the momentum window coincides with that in the flow equations; however, in the former integral, significant deviations from the LPA' are expected in the UV. Indeed, it has been shown in \cite{Lahoche:2018oeo} that using $f$ as given by the truncation to evaluate an unbounded UV integral leads to paradoxes.\\

Taking the third functional derivative with respect to the classical field and once with respect to the response field, we obtain a relation linking the variation of the 4-point vertex function with respect to the external momenta to the local quartic and sextic couplings. Graphically:
\begin{align}
\nonumber&-\vcenter{\hbox{\includegraphics[scale=0.7]{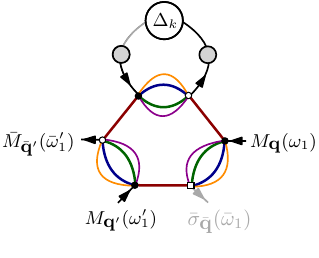}}}-\vcenter{\hbox{\includegraphics[scale=0.7]{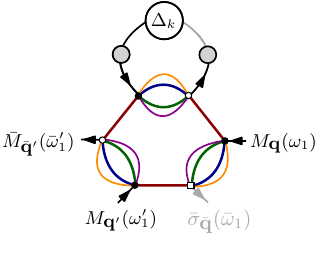}}}\\\nonumber
&+\vcenter{\hbox{\includegraphics[scale=0.7]{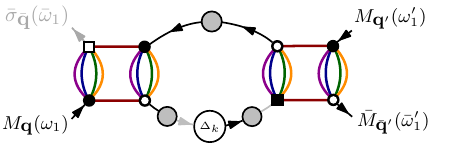}}}+\vcenter{\hbox{\includegraphics[scale=0.7]{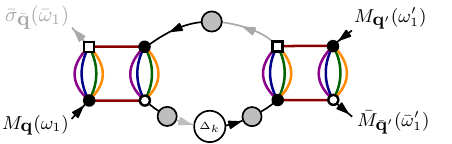}}}\\\nonumber
&+\vcenter{\hbox{\includegraphics[scale=0.7]{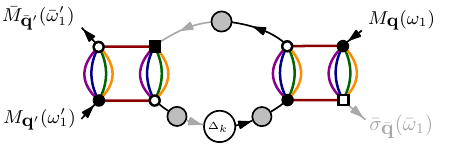}}}+\vcenter{\hbox{\includegraphics[scale=0.7]{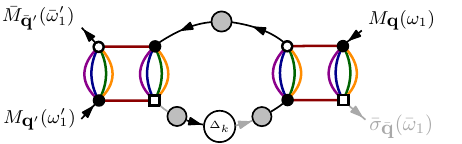}}}\\\nonumber
&+\vcenter{\hbox{\includegraphics[scale=0.7]{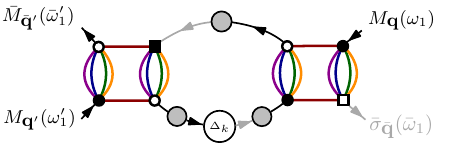}}}+\vcenter{\hbox{\includegraphics[scale=0.7]{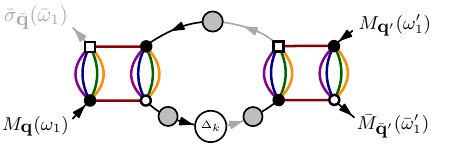}}}\\
&-\left(\vcenter{\hbox{\includegraphics[scale=0.7]{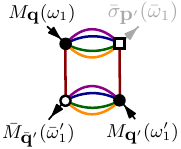}}}+\vcenter{\hbox{\includegraphics[scale=0.7]{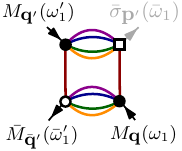}}}\right)\delta_{\bm{p}\bar{\bm{q}}}+\left(\vcenter{\hbox{\includegraphics[scale=0.7]{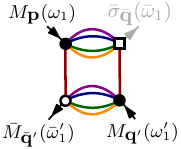}}}+\vcenter{\hbox{\includegraphics[scale=0.7]{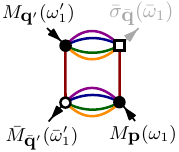}}}\right)\delta_{\bm{p}^\prime \bm{q}}=0\,.
\end{align}\label{relationWard2}
Translating this relation into equation, and using \eqref{Wardcontinuum}, we get:
\begin{align}
\bar{\lambda}^\prime= \frac{32\bar{\lambda}^2}{3\pi} \int_{\mathbb{R}^4} d\textbf{x} \int dy\frac{1}{f^3(x,y)f(x,-y)}+\frac{8\bar{\lambda}^2}{\pi} \left(\mathcal{\bar{L}}_{k,2}^{(1)}\vert_{Z_\infty=0}+\mathcal{\bar{L}}_{k,2}^{(2)}\vert_{Z_\infty=0} \right)\,,
\end{align}
where:
\begin{align}
\nonumber\mathcal{\bar{L}}_{k,1}:=2\int dy \int_{\mathbb{R}^4} d\bm x \bigg[&\left(Z_{\infty}{Z}^{-1}(k)- \theta(1-x)\right)\frac{1}{\hat{f}(x,y)\hat{f}^2(x,-y)}\bigg]\,,
\end{align}
\begin{align}
\nonumber \mathcal{\bar{L}}_{k,2}^{(1)}:= -\int dy \int_{\mathbb{R}^4} d\bm x \bigg[&\left(Z_{\infty}{Z}^{-1}(k)-  \theta(1-x)\right)\frac{1}{\hat{f}(x,y)\hat{f}^2(x,-y)}\bigg]\,.
\end{align}

The resulting flow equations for $m^2$ and $\lambda$ formally resemble those obtained for the equilibrium theory in \cite{Lahoche_2020b}, which is not surprising given the assumption of equilibrium dynamics. However, although the equations are the same, their interpretation differs (see below). In particular, we recover the fact that the theory is asymptotically free:
\begin{equation}
\dot{\bar{\lambda}} = - 4 \pi^2 \bar{\lambda}^2+\mathcal{O}(\bar{\lambda}^3)\,.\label{asymptoticfree}
\end{equation}
Moreover, no reliable global fixed point exists in the UV regime: the theory remains essentially trivial for much of its history. Let us mention that this situation could be different in the deep IR \cite{Benedetti_2016,Benedetti_2015,Geloun_2016}, although investigations beyond the symmetric-phase analysis seem to suggest a negative answer \cite{pithis2021no}. \\

\begin{figure}
\begin{center}
\includegraphics[scale=0.55]{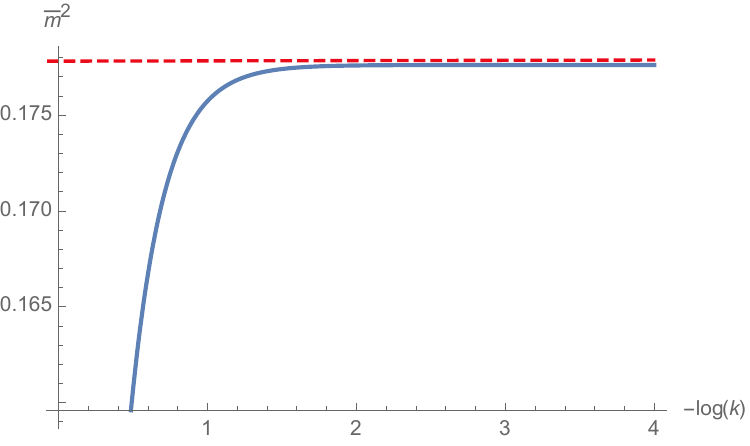}\qquad \includegraphics[scale=0.55]{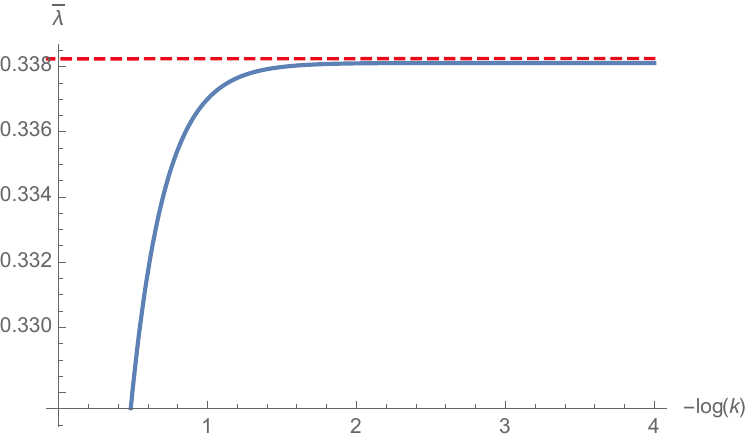}
\end{center}
\caption{Typical behavior of the RG flow in the vicinity of the Gaussian fixed point, initial conditions being for $\bar{m}^2(0)= -0.08$ and $\bar{\lambda}=0.1$.}\label{figdiv1}
\end{figure}

The absence of a reliable UV fixed point does not imply that the physics of the flow is trivial in the IR, as is well known for the non-perturbative nature of the IR limit of Yang-Mills theories \cite{weinberg1995quantum}. Indeed, by integrating the flow equation \eqref{asymptoticfree}, we find, for instance:

\begin{equation}
\lambda(k)=\frac{\lambda_0}{1+4\pi^2 \ln(k/\Lambda)}\,,
\end{equation}
where $\lambda_0 := \lambda(\Lambda)$ is the UV coupling. A finite RG-time singularity occurs at $k = \Lambda e^{\frac{1}{4\pi^2 \lambda_0}}$. Numerical integration of the nonperturbative flow shows that this pole, which notably ignores mass feedback, is essentially a perturbative artifact. As shown in Figure \ref{figdiv1}, the flow converges after a transient regime. However, this conclusion is partly incorrect, as it is linked to the presence of a "false" attractive fixed point. A more elaborate method, which we will develop later, will eliminate this problem and indicate that the phenomenon of a singularity at a finite scale is likely unavoidable. We will return to this issue in Section \ref{numan}.

\section{Exploring dynamics beyond equilibrium}

The method developed in \cite{lahoche2023stochastic}, which we have summarized, applies under the assumption of equilibrium dynamics; that is, for a system that starts in equilibrium at $t = -\infty$ and returns to it at $t = +\infty$. In this regime, the evolution is both time-reversal invariant (as expressed by the FDT) and invariant under time translations. Here, we consider a parameterization that allows deviations from time-reversal symmetry. The procedure we propose has been previously applied to disordered random matrices by the same authors in \cite{lahoche2024functional}, and the discussion below is inspired by that reference. \\

\subsection{Parametrization}

The fluctuation-dissipation theorem \eqref{FPTF} comes from time reversal symmetry \cite{aron2010symmetries2}, and is formally a direct consequence of the following matricial form for the $\Gamma^{(2)}_k$ kinetic function:
 \begin{equation}
\Gamma^{(2)}_k(\bm{p},\omega) = \begin{pmatrix}
A& B(\bm{p},\omega) \\
B(\bm{p},-\omega)& 0
\end{pmatrix}
\,.
 \end{equation}
The time-reversal symmetry implies that the wave function renormalization for the response field equals the wave function renormalization for $\bar{\eta} \dot{T}$. In other words, if $(\Gamma^{(2)}k)_{\bar{\eta} \eta}=A$, then we must have $B(\bm{p},\omega)=- A \omega + i f_k(\bm{p})$. This simple observation suggests a way to move away from the FDT: one may choose a truncation that does not enforce this constraint (while still using a regulator that preserves it). For example, taking $B(\bm{p},\omega)=-\omega + i f_k(\bm{p},)$ but allowing $A\neq 1$. In this case, $A$ necessarily develops a non-trivial flow, which requires modifying the truncation so that $A$ can evolve, including interactions that would normally be forbidden by perturbation theory.

Such a situation arises, for instance, when the flow exhibits a finite-time singularity \cite{gredat2014finite,lahoche2025large,achitouv2024time,achitouv2025constructing}, as is expected for asymptotically free theories. This approach is not unusual: in the study of phase transitions (particularly first order, see \cite{lahoche2025large} and references therein), it is standard to assume that operators forbidden in perturbative expansions can become relevant in certain regions of phase space connected to the Gaussian fixed point, where they then play a central role in determining the IR physics (see also \cite{gredat2014finite,lahoche2025large,achitouv2024time,achitouv2025constructing}).
The role of these new interactions will be made clear in the numerical analysis of the flow equations, to which we will return in Section \ref{numan}.

\begin{figure}
\begin{center}
\includegraphics[scale=1.2]{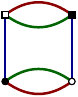} \qquad \includegraphics[scale=1.2]{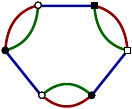} \qquad \includegraphics[scale=1.2]{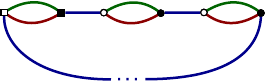}
\end{center}
\caption{The non-branching melons involving two response fields.}\label{figtworesponseexample}
\end{figure}

We will therefore restrict our attention to "new" interactions of the type shown in Figure \ref{figtworesponseexample}, namely cyclic melons with two conjugate response fields (see \cite{lahoche2024functional} for details). The quartic melons, for example, contribute both to the vertex $\bar{\eta}\eta$ and to corrections of the two-point vertex $\bar{\eta}\Phi$. As a result, the flows of mass and coupling $(\Gamma^{(2)}_k)_{\bar{\eta} \eta}$ are not independent, while causality guarantees that the flow equation for $u_2$ remains essentially unaffected.

For the kinetic part of the effective average action, the truncation reads: 
\begin{equation}
\Gamma_k^{(2)}(\bm p,\omega)=\begin{pmatrix}
1+\Delta +  \Delta^\prime\, \bm p^2 & -\omega+iZ(k)(\bm p^2+k^2\bar{m}^2) \\
\omega+iZ(k)(\bm p^2+k^2\bar{m}^2)& 0
\end{pmatrix}
\,.\label{matrixGamma2}
\end{equation}
Note that the contribution $ \Delta^\prime\, \bm p^2$ arises because the melonic quartic vertices with two response fields have dimension $-2$, the same as $ \Delta^\prime$.\\

In the next two sections, we will explore this parametrization using two types of approximations. The first is a semi-perturbative expansion, in which interactions normally forbidden by perturbation theory are treated perturbatively via a vertex expansion. The second is an improved scheme that relies on the machinery of Ward identities. For clarity, we will from now on refer to interactions with a single response field as monogamous interactions, and to those involving two combined response fields as bigamous interactions.

\subsection{Breaking FDT: Expansion around equilibrium dynamics}

In this first part, we develop a simple, easy-to-implement approximation, which consists essentially in retaining only the first-order feedback of bigamous interactions on monogamous ones. Let us briefly recall the general philosophy. In this region of phase space, we assume that instabilities arise which generate operators breaking time-reversal symmetry. Our aim is to determine whether such operators become relevant in the large-scale description of the theory. Since the theory is constructed as the asymptotic one at $t=+\infty$, the corresponding IR regime is defined by small momenta $\bm{p}$.
Because these interactions are forbidden by perturbation theory, standard melonic relations between six- and four-point functions—built from formal resummations of the (renormalized) perturbative series—lack contributions from bigamous interactions. Denoting by $\lambda_2$ and $\kappa_2$ the melonic bigamous interactions represented in Figure \ref{figtworesponseexample}, the contribution of bigamous interactions to $\kappa$ is expected to be, at best, of order $\mathcal{O}(\lambda^2 \lambda_2)$, which sets the limit of validity of the approximation.
Finally, we will neglect the momentum dependence of the bigamous vertices, a simplification that facilitates the computation of the flow of $\Delta^\prime$. The discarded contributions are expected to be of order $\lambda_2^2$. \\

Let us start with the flow of the $\bar{\sigma} M$ component of the two-point function. According to the general method described in Section \ref{method}, and using equations \eqref{eqflowMass}, a direct inspection shows that no melonic contributions arise from bigamous interactions. Consequently, the graphical flow equation remains unchanged from the previous case, namely:
\begin{align}
\nonumber\dot{\gamma}_{k,\bar{\sigma}M}^{(2)}(\bm{p}_1,\hat{\omega}_1)\delta_{\bm{p}_1\bm{p}_2}\delta(\hat{\omega}_1-\hat{\omega}_2)&=-\sum_{i=1}^d\,\vcenter{\hbox{\includegraphics[scale=1]{OneLoopMass2.pdf}}}\,.\label{flowmassdiag}
\end{align}

Hence, the $\beta$-function for the mass reads, computing the integrals in \eqref{betam0} and taking into account the contributions arising because of $\Delta$ and $\Delta^\prime$ in $G_{\bar{M} M}$:

\begin{equation}
\boxed{\dot{\bar{m}}^2=-(2+\eta)\bar{m}^{2}-10 \pi^2 \bar{\lambda}\,\frac{1+\Delta+2 \bar{\Delta}^\prime/3}{(1+\bar{m}^{2})^2}\,\left(1+\frac{\eta}{6}\right)\,.}\label{betamNew}
\end{equation}

Moreover, $\eta$ is again given by equation \eqref{eqeta}, and we will return on the computation of $\lambda^\prime$ below. Now, let us move on the computation of the flow equations for $\Delta$ and ${\Delta}^\prime$. We get:

\begin{equation}
(\dot{\Delta}+\dot{\Delta}^\prime\, \bm p^2_1)\,\delta_{\bm{p}_1\bm{p}_2}\delta(\hat{\omega}_1-\hat{\omega}_2)= -\sum_{i=1}^d\,\vcenter{\hbox{\includegraphics[scale=1]{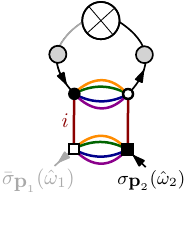}}}\,.\label{flowmassdiag}
\end{equation}
The loop integral is the same as for the mass, but $\Delta$ is dimensionless and $\lambda_2$ must have dimension $-2$, i.e. $\lambda_2:= k^{-2} \bar{\lambda}_2$. We then have, setting external momenta to zero\footnote{Note that the coupling $\lambda_2$ arises with a factor $-i$ to agree with the formula above.}:

\begin{equation}
\boxed{\dot{\Delta}=- 10\pi^2\bar{\lambda}_2\,\frac{1+\Delta+2 \bar{\Delta}^\prime/3}{(1+\bar{m}^{2})^2}\,\left(1+\frac{\eta}{6}\right)\,.}\label{eqDelta}
\end{equation}

Note that we choose the same numerical conventions for the normalization of bigamous and monogamous interactions, explaining the factor $10=2\times 5$. The extraction of $\Delta^\prime$ follows the same strategy as for the anomalous dimension (neglecting the full momentum dependency of the vertex here):
\begin{equation}
\dot{\Delta}^\prime=-{\lambda}_2 \frac{d}{d p_1^2} L_{21}(x_1)\big\vert_{x_1=0}\,.\label{deltaprime}
\end{equation}
The integral can be easily computed exactly, and expanding it at the leading order in $p^2_1$, leads to:
\begin{align}
k^2\dot{\Delta}^\prime=-2 \pi^2 \bar{\lambda}_2 \frac{\bar{\Delta}^\prime}{(1+\bar{m}^{2})^2}\,\left(1+\frac{\eta}{6}\right)+2 \pi^2  {\lambda}_2 \frac{\eta(1+\bar{\Delta})}{(1+\bar{m}^{2})^2}+\frac{4 \pi^2}{3} \bar{\lambda}_2 \frac{1+\bar{\Delta}+\bar{\Delta}^\prime}{(1+\bar{m}^{2})^2}\,,
\end{align}
or, by arranging a little:
\begin{equation}
\boxed{\dot{\bar{\Delta}}^\prime=2\bar{\Delta}^\prime-\frac{2 \pi^2 \bar{\lambda}_2}{(1+\bar{m}^{2})^2} \left( \frac{2}{3}\left(1+\bar{\Delta}-\frac{\bar{\Delta}^\prime}{2}\right)+ \eta \left(1+\bar{\Delta}-\frac{\bar{\Delta}^\prime}{6}\right)\right)\,.\label{eqDeltaprime}}
\end{equation}

The expression for $\eta$ has been recalled in the section \ref{method}, equation \eqref{}. We have, explicitly:

\begin{equation}
\eta=-2\bar{\lambda}^\prime \pi^2 \,\frac{1+\bar{\Delta}+2 \bar{\Delta}^\prime/3}{(1+\bar{m}^{2})^2}\,\left(1+\frac{\eta}{6}\right)+\bar{\lambda} \pi^2\frac{1+\bar{\Delta}+2 \bar{\Delta}^\prime/3}{(1+\bar{m}^{2})^2} \left(4+\eta\right)- \frac{2 \pi^2 \bar{\lambda}\bar{\Delta}^\prime}{(1+\bar{m}^{2})^2}\left(1+\frac{\eta}{6}\right)\,,
\end{equation}

which can be solved as:

\begin{equation}
\boxed{\eta=2 \pi^2\frac{(2\bar{\lambda}-\bar{\lambda}^\prime)(1+\bar{\Delta}+2 \bar{\Delta}^\prime/3)-\bar{\lambda}\bar{\Delta}^\prime}{(1+\bar{m}^{2})^2+(\bar{\lambda}^\prime-3\bar{\lambda}) \pi^2 (1+\bar{\Delta}+2 \bar{\Delta}^\prime/3)/3+\pi^2 \bar{\lambda} \bar{\Delta}^\prime/3}\,.}\label{etafinal}
\end{equation}

We will restrict the truncation to sextic order for monogamous interactions and to quartic order for bigamous interactions, a choice consistent with power-counting arguments. In principle, the approximation could be refined by a systematic analysis of higher-order contributions and by examining the possible convergence of IR properties, such as the critical exponents associated with potential fixed points. Such investigations have been carried out, for instance, in \cite{Carrozza_2017a}. Nevertheless, this line of study will ultimately be superseded by the more exact method presented in the following section, of which the present discussion should be viewed as an introduction.

Let us first turn to the flow equation for $\lambda$. We make the simplifying assumption that the monogamous sextic interaction $\kappa$ remains unaffected by the presence of bigamous interactions. In other words, we assume that the exact relations inherited from perturbation theory remain valid, up to the point where they would break down, the neglected terms being at most of order $\lambda^2 \lambda_2$. As a result, part of the $\beta$-function for $\lambda$ has, in principle, already been computed. It suffices to replace $1 \to 1+\Delta+2\Delta^\prime/3$ in the numerators of the effective loops, as illustrated in the computation of the mass flow. We denote this contribution by $\dot{\bar{\lambda}}\vert_\text{mono}$.

\begin{align}
\nonumber\dot{\bar{\lambda}}\vert_\text{mono}\approx-2\eta \bar{\lambda}+&4\pi^2\bar{\lambda}^2 \,\frac{1+\Delta+2 \bar{\Delta}^\prime/3}{(1+\bar{m}^{2})^3}\,\left(1+\frac{\eta}{6}\right)\Big[1\\
&-6\pi^2\bar{\lambda}\left(\frac{1}{(1+\bar{m}^{2})^2}+\left(1+\frac{1}{1+\bar{m}^{2}}\right)\right)\Big]\,.
\end{align}

\begin{figure}
\begin{center}
\includegraphics[scale=1]{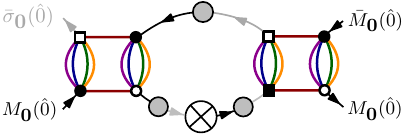}\quad \includegraphics[scale=1]{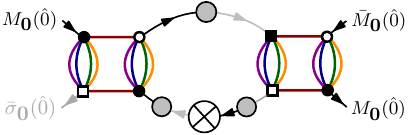} 
\end{center}
\caption{Contributions for the contribution of bigamous vertices to $\dot{\bar{\lambda}}\vert_\text{big}$.}\label{figlambdalambda2}
\end{figure}

The flow still needs to be completed by the contribution of the bigamous melons. It is easy to see that there can be no melonic contributions of the type $\lambda_2^2$, and the only terms actually contributing to the flow of $\lambda$ will be of the form $\lambda \lambda_2$, as shown in Figure \ref{figlambdalambda2}. Let us call this contribution $\dot{\bar{\lambda}}\vert_\text{big}$, we have, computing the effective loop:

\begin{equation}
\dot{\bar{\lambda}}\vert_\text{big}=4\pi^2 \,\frac{\bar{\lambda}\bar{\lambda}_2}{(1+\bar{m}^{2})^2}\,\left(1+\frac{\eta}{6}\right)\,,
\end{equation}
then, we obtain the final approximation for $\dot{\bar{\lambda}}$:

\begin{align}
\nonumber\dot{\bar{\lambda}}\approx-2\eta \bar{\lambda}+4\pi^2\bar{\lambda}^2 \,\frac{1+\Delta+\frac{2}{3} \bar{\Delta}^\prime}{(1+\bar{m}^{2})^3}&\,\left(1+\frac{\eta}{6}\right)\Big[1+\frac{ \bar{\lambda}_2}{\bar{\lambda}} \frac{1+\bar{m}^{2}}{1+\Delta+\frac{2}{3} \bar{\Delta}^\prime}\\
& -6\pi^2\bar{\lambda}\left(\frac{1}{(1+\bar{m}^{2})^2}+\left(1+\frac{1}{1+\bar{m}^{2}}\right)\right)\Big]\,.\label{eqlambda}
\end{align}

\begin{figure}
\begin{center}
\includegraphics[scale=1]{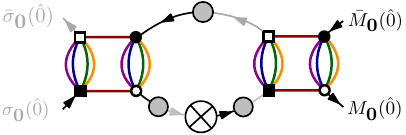}\quad \includegraphics[scale=1]{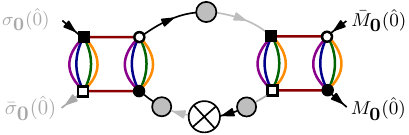}\\
\includegraphics[scale=1]{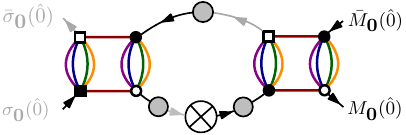}\quad \includegraphics[scale=1]{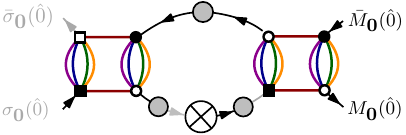} \\
\includegraphics[scale=1]{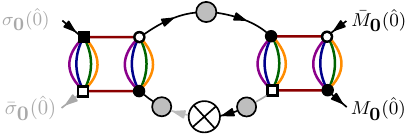}\quad \includegraphics[scale=1]{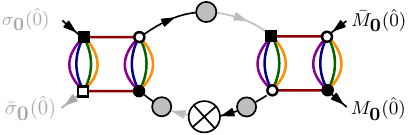} \\
\end{center}
\caption{Contributions for the contribution of bigamous vertices to $\dot{\bar{\lambda}}_2$.}\label{figlambda2dot}
\end{figure}

Now, we have to compute the last piece, $\dot{\bar{\lambda}}_2$. One can expect two kinds of contributions, like $\lambda \lambda_2$ and like $\lambda_2^2$, both pictured on Figure \ref{figlambda2dot}. Note that, as announced, we neglect the contributions of bigamous sextic interactions. The computations of the relevant diagrams follows the same method as before, and, taking into account symmetry factors, we get:

\begin{equation}
\boxed{\dot{\bar{\lambda}}_2=(2-\eta) \bar{\lambda}_2+ \pi^2\bar{\lambda}\bar{\lambda}_2 \,\frac{1+\Delta+\frac{2}{3} \bar{\Delta}^\prime}{(1+\bar{m}^{2})^3}\,\left(1+\frac{\eta}{6}\right)+ 2\pi^2 \,\frac{\bar{\lambda}_2^2}{(1+\bar{m}^{2})^2}\,\left(1+\frac{\eta}{6}\right)\,.}\label{eqlambda2}
\end{equation}

The numerical investigation of these equations is reported in section \ref{numan}. \\

\subsection{Ward identities method}\label{secWard}

In this section, we aim to improve upon the previous expansion by incorporating the constraints imposed by Ward identities, following the general framework of effective vertex expansion developed in \cite{Lahoche:2018oeo}. It is worth emphasizing once more that, although this strategy has not always been implemented in renormalization group analyses of TGFTs (primarily due to technical challenges associated with non-locality), it is by no means unusual. Indeed, Ward identities play a central role in ordinary quantum electrodynamics, where they enforce, for example, the equality between wave-function renormalization and vertex renormalization.

\begin{figure}
\begin{center}
\includegraphics[scale=1.2]{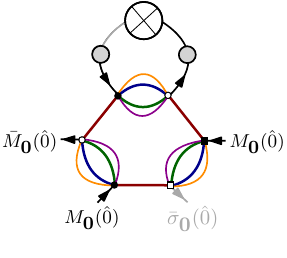}
\end{center}
\caption{Sextic ($\kappa_2$) contribution to the flow of  ${\bar{\lambda}}_2$.}\label{contphi6lambda2}
\end{figure}

As a first step, since it will be needed later, we establish the full flow equation for $\lambda_2$, including the sextic contributions. The corresponding diagrams are displayed in Figure \ref{contphi6lambda2}, and their evaluation proceeds along the same lines as the computation of the mass flow, differing only by a numerical factor. We thus obtain:

\begin{align}
\nonumber \dot{\bar{\lambda}}_2=&(2-\eta) \bar{\lambda}_2-3 \pi^2 \bar{\kappa}_2\,\frac{1+\Delta+2 \bar{\Delta}^\prime/3}{(1+\bar{m}^{2})^2}\,\left(1+\frac{\eta}{6}\right)\\
&+ 2\pi^2\bar{\lambda}\bar{\lambda}_2 \,\frac{1+\Delta+\frac{2}{3} \bar{\Delta}^\prime}{(1+\bar{m}^{2})^3}\,\left(1+\frac{\eta}{6}\right)+ 4\pi^2 \,\frac{\bar{\lambda}_2^2}{(1+\bar{m}^{2})^2}\,\left(1+\frac{\eta}{6}\right)\,.\label{eqlambda22}
\end{align}

We moreover recall the expression of $\dot{\bar{\lambda}}$ with explicit monogamous sextic coupling  $\kappa$:

\begin{align}
\dot{\bar{\lambda}}=-2\eta \bar{\lambda}-3 \pi^2 \bar{\kappa}\,\frac{1+\Delta+2 \bar{\Delta}^\prime/3}{(1+\bar{m}^{2})^2}\,\left(1+\frac{\eta}{6}\right)+4\pi^2\bar{\lambda}^2 \,\frac{1+\Delta+\frac{2}{3} \bar{\Delta}^\prime}{(1+\bar{m}^{2})^3}\,\left(1+\frac{\eta}{6}\right)\,.\label{eqlambdaKappa}
\end{align}

We will also need to improve our calculation of the flow of $\Delta^\prime$, taking into account, as we did for $\eta$, the vertex-specific moment dependence. We obtain by a similar calculation:

\begin{align}
\nonumber \dot{\bar{\Delta}}^\prime=2\bar{\Delta}- 2\pi^2&\bar{\lambda}_2^\prime\,\frac{1+\Delta+2 \bar{\Delta}^\prime/3}{(1+\bar{m}^{2})^2}\,\left(1+\frac{\eta}{6}\right)\\
&-\frac{2 \pi^2 \bar{\lambda}_2}{(1+\bar{m}^{2})^2} \left( \frac{2}{3}\left(1+\bar{\Delta}-\frac{\bar{\Delta}^\prime}{2}\right)+ \eta \left(1+\bar{\Delta}-\frac{\bar{\Delta}^\prime}{6}\right)\right)\,,\label{eqDeltaprime2}
\end{align}
where $\bar{\lambda}_2^\prime$ is the (dimensionless) derivative of the vertex with respect to the external momenta, analogous to \eqref{derivvertex}.\\

Now, let us focus on the Ward identities. Relevant relations concerned by the truncation we consider can be obtained from the statement \ref{Ward1}, taking derivatives $\partial^2/\partial M_{\bm q}({\omega}_1)\partial \bar{\sigma}_{\bm{\bar{q}}}({\bar{\omega}}_1)$ and $\partial^2/\partial \sigma_{\bm q}({\omega}_1)\partial \bar{\sigma}_{\bm{\bar{q}}}({\bar{\omega}}_1)$, setting classic field to zero at the end of the computation (symmetric phase condition). Let us investigate them separately. Graphically, the resulting relations for the first one leads to, keeping only the melonic contributions:

\begin{align}
\nonumber &\Bigg(\vcenter{\hbox{\includegraphics[scale=0.8]{OneLoopMassWI1BBB.pdf}}}+\vcenter{\hbox{\includegraphics[scale=0.8]{OneLoopMassWI4BBB.pdf}}}\Bigg)\delta_{p^\prime q_i} \delta_{p\bar{q}_i}\prod_{j\neq i}\delta_{q_j{\bar{q}}_j}\delta(\omega_1-\bar{\omega}_1)\\\nonumber
&-\sum_{\bm p, \bm p^\prime} \prod_{j\neq i} \delta_{p_jp_j^\prime} \Delta_k(\bm p,\omega_1) \delta_{\bm p\bm{\bar{q}}} \delta_{\bm q\bm p^\prime}\delta(\omega_1-\bar{\omega}_1) +\sum_{\bm p, \bm p^\prime} \prod_{j\neq i} \delta_{p_jp_j^\prime}\bigg[\gamma_{k,\bar{\sigma}M}^{(2)}(\bm p,\omega_1)- \gamma_{k,\bar{\sigma}M}^{(2)}(\bm p^\prime,\omega_1)
\bigg] \\
&\times \delta_{\bm p^\prime \bm q}\delta_{\bm p\bm{\bar{q}}} \delta_{p_ip}\delta_{p_i^\prime p^\prime}\delta(\omega_1-\bar{\omega}_1)+ \sum_{\bm p, \bm p^\prime} \prod_{j\neq i} \delta_{p_jp_j^\prime}\delta_{\bm p\bm{\bar{q}}}\delta_{\bm q\bm p^\prime} i\delta R_k^{(1)}(\bm p,\omega_1)\delta(\omega_1-\bar{\omega}_1)=0\,,
\end{align}
where the diagonal operator $\Delta_k$ is defined by \eqref{deltadef}. Note that bigamous vertices are absent, and it is easy to check that their contributions involve the propagator $G_{\bar{\chi}\chi}$, which vanishes. Thus, formally, the relation is the same as in the equilibrium dynamics phase and simplifies to:
\begin{equation}
\vcenter{\hbox{\includegraphics[scale=0.8]{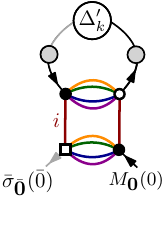}}}+\vcenter{\hbox{\includegraphics[scale=0.8]{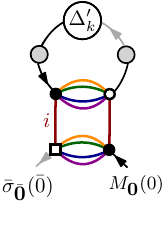}}}+i\left(Z-Z_{\infty}\right)=0\,,
\end{equation}
where the following approximation is employed (see \cite{lahoche2023stochastic} for details):
\begin{equation}
\big[\gamma_{k,\bar{\sigma}M}^{(2)}(\bm p,0)- \gamma_{k,\bar{\sigma}M}^{(2)}(\bm p^\prime,0)\big]\big\vert_{p_i=0} \approx iZ \delta p^2 + \mathcal{O}(\delta p^2)\,.\label{equationde}
\end{equation}

Formally, the resulting relation retains the same structure as at equilibrium, with the difference that, in the "IR" interval selected by the regulator derivative in $\Delta_k^\prime$—which coincides with the momentum window selected by $\dot{R}_k$ in the flow equation—the propagator must be replaced by its truncated form \eqref{matrixGamma2} (see \cite{Lahoche:2018oeo,Lahoche_2020b,Lahoche_2019bb} for an extended discussion). Explicitly:
\begin{equation}
\boxed{2 \lambda(k) \mathcal{L}(k)=Z_\infty-Z(k)\,,}\label{Wardidentity01}
\end{equation}
where $\lambda(k)$ is the renormalized quartic coupling and:

\begin{equation}
\mathcal{L}(k):=\mathcal{L}^{(\infty)}(k)+\mathcal{L}^{(0}(k)\,,
\end{equation}
where:
\begin{equation}
\mathcal{L}^{(0)}(k):= k^2 Z(k)\sum_{\bm p \in \mathbb{Z}^4} \frac{\partial }{\partial p_1^2}r_k(\bm p)  \frac{1+\Delta+\Delta^\prime \bm p^2}{(Z \bm p^2+m^2+Z k^2 r_k(\bm p^2))^2}\,,\label{L0}
\end{equation}
and,
\begin{equation}
\mathcal{L}^{(\infty)}(k):=\frac{4iZ_{\infty}\lambda(k)}{\pi}\int d\omega  \sum_{\bm p \in \mathbb{Z}^{4}} \, G_{k,\bar{M} M}(\bm p,\omega) G_{k,\bar{\sigma} M}(\bm p,\omega)\,.\label{Linfty}
\end{equation}
This expression is admittedly obscure, and as emphasized earlier, one cannot simply replace the propagator components by their truncation; doing so leads to inconsistencies arising from the global UV divergence, as analyzed in \cite{Lahoche:2018oeo}. Nevertheless, we shall see that this calculation is, to some extent, unnecessary if a few assumptions are accepted. The first of these is that $Z_\infty$ is essentially determined by the equilibrium theory (see, for instance, \cite{Lahoche_2020b} and the appendix of the preceding work \cite{lahoche2023stochastic}):
\begin{equation}
Z_\infty= \frac{1}{1-2 \lambda_r \mathcal{A}_2}\,,\label{Zinfty}
\end{equation}
where $\lambda_r$ is the renormalized constant (for the IR equilibrium theory, the vertex renormalization equaling the wave function renormalization), and:

\begin{equation}
\mathcal{A}_2=\sum_{\bm p \in \mathcal{Z}^4} \, G_k^2(\bm p^2)\,,
\end{equation}
$G_k$ is the renormalized propagator of the equilibrium theory. Moreover, let us recall that the exact $\beta$-function for the renormalized coupling $\lambda(k)$ 

\begin{equation}
\dot{\lambda}(k)=-2 \lambda^2(k) \, \dot{\mathcal{A}}_2\,.\label{eqexactlambda}
\end{equation}
All these results can be found in \cite{Lahoche_2020b} and the references therein; we recall them here for the discussion. For instance, the result \eqref{eqexactlambda} shows that, since $\lambda(k)$ denotes the renormalized coupling at the scale $k$, it must be finite, and consequently $\dot{\mathcal{A}}_2$ must also be convergent. This expectation arises when exchanging the order of differentiation and integration, and observing that the regulator affects only the differences in the region where it contributes significantly. The argument is, however, somewhat subtle; see, for example, the discussion in \cite{Cotler_2023}. In this work, we prefer to rely on the interpretation of \eqref{eqexactlambda}. Furthermore, recall from \eqref{Zinfty} that the continuum limit imposes $Z_\infty^{-1} \sim \ln \Lambda$, as already noted.\\

Let us compute the derivative of \eqref{Wardidentity01} with respect to $\ln k/\Lambda$, we get:

\begin{equation}
2 \lambda \mathcal{L}(k)\frac{\dot{\lambda}}{\lambda}+2 \lambda \dot{\mathcal{L}}(k)=- Z \eta\,. 
\end{equation}
The first term can be replaced by $-Z(k) \approx 2 \lambda \mathcal{L}(k)$ (in the continuum limit). Moreover,
\begin{equation}
\dot{\mathcal{L}}(k)=\dot{\mathcal{L}}^{(\infty)}(k)+ \dot{\mathcal{L}}^{(0)}(k)\,.
\end{equation}
The second term of this expression is straightforward to compute from \eqref{L0}. The first term, however, involves the derivative with respect to $\ln (k/\Lambda)$ of a quantity that vanishes in the equilibrium dynamics. We further expect that the modifications induced by the bigamous operators affect this quantity only in the IR, without altering the ultraviolet contributions, which, as shown by \eqref{eqexactlambda}, must be suppressed through multiplication by $Z_\infty$. On this basis, we adopt the following reasonable hypothesis:
\begin{equation}
\dot{\mathcal{L}}^{(\infty)}(k)\approx 0\,.
\end{equation}
Then, because
\begin{equation}
\frac{\partial }{\partial p_1^2}r_k(\bm p) = - \frac{1}{k^2} \theta (k^2-\bm p^2)\,,
\end{equation}
we find:
\begin{equation}
\mathcal{L}^{(0)}(k)= -\frac{\pi^2}{2Z}\frac{1+\bar{\Delta}+\frac{2}{3}\bar{\Delta}^\prime}{(1+\bar{m}^2)^2}\,.
\end{equation}
and derivation with respect to $\ln(k/\Lambda)$ leads to:
\begin{equation}
\dot{\mathcal{L}}^{(0)}(k)=-\frac{\pi^2}{2 Z^2}\frac{1+\bar{\Delta}+\frac{2}{3}\bar{\Delta}^\prime}{(1+\bar{m}^2)^2} \left(-\eta+\frac{\dot{\bar{\Delta}}+\frac{2}{3}\dot{\bar{\Delta}}^\prime}{1+\bar{\Delta}+\frac{2}{3}\bar{\Delta}^\prime}-\frac{2 \dot{\bar{m}}^2}{1+\bar{m}^2}\right)\,.
\end{equation}
Now, taking into account that, from definition:

\begin{equation}
\dot{\bar{\lambda}}= \frac{\dot{\lambda}-2\eta \lambda}{Z^2}\,,
\end{equation}
we have the constraint:
\begin{equation}
\boxed{\dot{\bar{\lambda}}+ \eta \bar{\lambda}-\bar{\lambda}^2\pi^2\frac{1+\bar{\Delta}+\frac{2}{3}\bar{\Delta}^\prime}{(1+\bar{m}^2)^2} \left(\eta-\frac{\dot{\bar{\Delta}}+\frac{2}{3}\dot{\bar{\Delta}}^\prime}{1+\bar{\Delta}+\frac{2}{3}\bar{\Delta}^\prime}+\frac{2 \dot{\bar{m}}^2}{1+\bar{m}^2}\right)=0\,.}\label{wardcont1}
\end{equation}
This relation coincides with the one obtained in \cite{Lahoche:2018oeo}, up to the modifications introduced by the bigamous operators. It is worth noting that, in this context, these operators appear as a natural extension of equilibrium dynamics, yet they seem to arise only within the stochastic formalism.\\

Now consider the relation coming from the derivative of the general Ward identity \eqref{Ward1} by $\partial^2/\partial \sigma_{\bm q}({\omega}_1)\partial \bar{\sigma}_{\bm{\bar{q}}}({\bar{\omega}}_1)$. Because in the symmetric phase\footnote{We drop field indices here to simplify notations.}:
\begin{equation}
\frac{\partial^2}{\partial \sigma\partial \bar{\sigma}} G_{k,\bar{\sigma} M} = -G_{k,\bar{\sigma} \bullet} \Gamma^{(4)}_{\sigma\bar{\sigma} \bullet \bullet} G_{k,\bullet M}\,,
\end{equation}
where the $\bullet$ materialize fields, and the construction of the relation have to be understood as a matrix product. The only suitable configuration is the following:
\begin{equation}
\frac{\partial^2}{\partial \sigma\partial \bar{\sigma}} G_{k,\bar{\sigma} M} = -G_{k,\bar{\sigma} M} \Gamma^{(4)}_{\sigma\bar{\sigma} M \bar{M}} G_{k,\bar{M} M}\,.
\end{equation}
Then, graphically, and in the melonic regime, the relation reads:
\begin{equation}
\vcenter{\hbox{\includegraphics[scale=0.8]{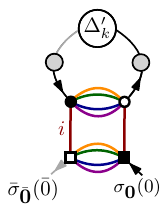}}}+\vcenter{\hbox{\includegraphics[scale=0.8]{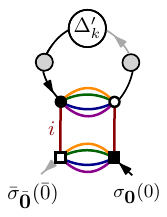}}}+ \Delta^\prime=0\,.
\end{equation}
The loop integral are the same as before, and the same arguments can be repeated here. We get:
\begin{equation}
2 \lambda_2 \mathcal{L}+\Delta^\prime=0\,. 
\end{equation}
Taking the first derivative with respect to $\ln(k/\Lambda)$, the relation simplifies as:
\begin{equation}
- Z(k) \frac{\dot{\lambda}_2}{\lambda} + 2 \lambda_2 \dot{\mathcal{L}}^{(0)}+\dot{\Delta}^\prime=0\,,
\end{equation}
or, again,
\begin{equation}
\boxed{\dot{\bar{\lambda}}_2-(2-\eta) \bar{\lambda}_2 -  \bar{\lambda}_2 \bar{\lambda} \pi^2\frac{1+\bar{\Delta}+\frac{2}{3}\bar{\Delta}^\prime}{(1+\bar{m}^2)^2} \left(\eta-\frac{\dot{\bar{\Delta}}+\frac{2}{3}\dot{\bar{\Delta}}^\prime}{1+\bar{\Delta}+\frac{2}{3}\bar{\Delta}^\prime}+\frac{2 \dot{\bar{m}}^2}{1+\bar{m}^2}\right)-\bar{\lambda}(\dot{\bar{\Delta}}-2 \bar{\Delta})=0\,.}\label{wardcont2}
\end{equation}
From the equations \eqref{wardcont2} and \eqref{wardcont1}, associated with the flow equations, \eqref{eqlambdaKappa}, and \eqref{eqlambda22}, we deduce explicit expressions for $\kappa$ and $\kappa_2$. Let us call these two solutions $\kappa_{\text{dyn}}:=\kappa_{\text{dyn}}(m,\lambda,\lambda_2,\Delta,\Delta^\prime)$ and $\kappa_{2{\text{dyn}}}:=\kappa_{2\text{dyn}}(m,\lambda,\lambda_2,\Delta,\Delta^\prime)$ respectively. Now let's look at the two missing pieces, $\lambda^\prime$ and $\lambda_2^\prime$, for which we will need the expressions for $\kappa$ and $\kappa_2$.

\begin{remark}
Note that the method developed in \cite{Lahoche_2020b} has the clear advantage of avoiding the estimation of sextic couplings through the evaluation of an unbounded loop integral, as was required in the approach summarized in Section \ref{summary}. Nevertheless, both methods yield similar results at equilibrium, and in particular confirm the absence of reliable fixed points.
\end{remark}

The derivation of these relations follows the same method as in the equilibrium case, and the monogamous part retains the same structure as \eqref{relationWard2}. Taking two derivatives with respect to $M$, one with respect to $\bar{M}$, and one with respect to $\bar{\sigma}$, we obtain:

\begin{align}
\nonumber&-\vcenter{\hbox{\includegraphics[scale=0.7]{OneLoop6PtsWard1.pdf}}}-\vcenter{\hbox{\includegraphics[scale=0.7]{OneLoop6PtsWard2.pdf}}}\\\nonumber
&+\vcenter{\hbox{\includegraphics[scale=0.7]{OneLoop4Pts2Ward1.pdf}}}+\vcenter{\hbox{\includegraphics[scale=0.7]{OneLoop4Pts2Ward2.pdf}}}\\\nonumber
&+\vcenter{\hbox{\includegraphics[scale=0.7]{OneLoop4Pts2Ward3.pdf}}}+\vcenter{\hbox{\includegraphics[scale=0.7]{OneLoop4Pts2Ward4.pdf}}}\\\nonumber
&+\vcenter{\hbox{\includegraphics[scale=0.7]{OneLoop4Pts2Ward5.pdf}}}+\vcenter{\hbox{\includegraphics[scale=0.7]{OneLoop4Pts2Ward6.pdf}}}\\\nonumber
&+\vcenter{\hbox{\includegraphics[scale=0.7]{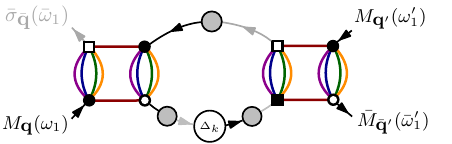}}}+\vcenter{\hbox{\includegraphics[scale=0.7]{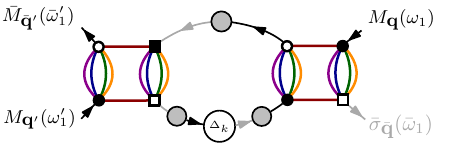}}}\\
&=\left(\vcenter{\hbox{\includegraphics[scale=0.7]{FourpointWard1.pdf}}}+\vcenter{\hbox{\includegraphics[scale=0.7]{FourpointWard2.pdf}}}\right)\delta_{\bm{p}\bar{\bm{q}}}-\left(\vcenter{\hbox{\includegraphics[scale=0.7]{FourpointWard3.pdf}}}+\vcenter{\hbox{\includegraphics[scale=0.7]{FourpointWard4.pdf}}}\right)\delta_{\bm{p}^\prime \bm{q}}\,.
\end{align}\label{relationWard00}

We assumed $\bm p^\prime \neq \bm q^\prime$ to cancel the last term on the right hand side, proportional to $\delta_{\bm p^\prime \bm q^\prime}$. This relation can be translated as a differential equation for $\pi_k^{(2)}$ as follows. We set $p_j=p_j^\prime =q_j=\bar{q}_j= 0 \, \forall j\neq i$, $p_i=p_i^\prime+1$, $\bm{q}^\prime=\bar{\bm{q}}^\prime=\bm 0$, $p_i= \bar{q}_i$ and $p_i^\prime=q_i$. Within this configuration, we showed in \cite{lahoche2023stochastic} that the two remaining terms on the right hand side becomes:

\begin{equation}
\frac{d}{dp_i^2}\pi_k^{(2)}(0,p_i^2)\bigg\vert_{p_i=0}=\frac{1}{2}\frac{d}{dp_i^2}\pi_k^{(2)}(p_i^2,p_i^2)\bigg\vert_{p_i=0} \,. 
\end{equation}

In the same way, taking one derivative with respect to $M$, one with respect to $\bar{M}$, one with respect to $\sigma$ and finally one with respect to $\bar{\sigma}$, we get:

\begin{align}
\nonumber&-\vcenter{\hbox{\includegraphics[scale=0.7]{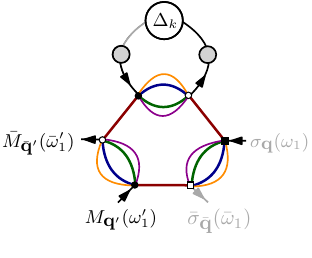}}}-\vcenter{\hbox{\includegraphics[scale=0.7]{OneLoop6PtsWard2.pdf}}}\\\nonumber
&+\vcenter{\hbox{\includegraphics[scale=0.7]{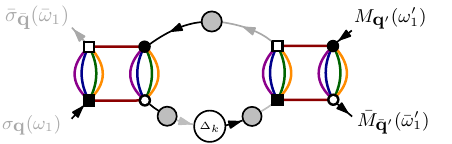}}}+\vcenter{\hbox{\includegraphics[scale=0.7]{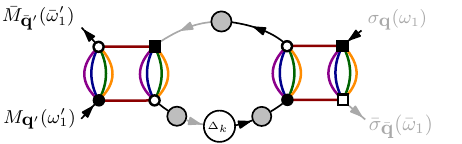}}}\\\nonumber
&+\vcenter{\hbox{\includegraphics[scale=0.7]{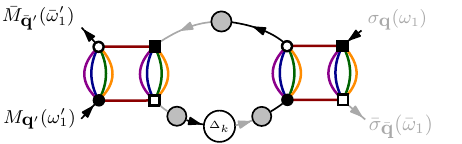}}}+\vcenter{\hbox{\includegraphics[scale=0.7]{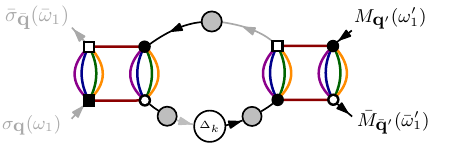}}}\\\nonumber
&+\vcenter{\hbox{\includegraphics[scale=0.7]{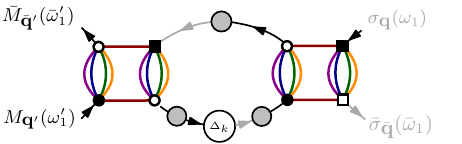}}}+\vcenter{\hbox{\includegraphics[scale=0.7]{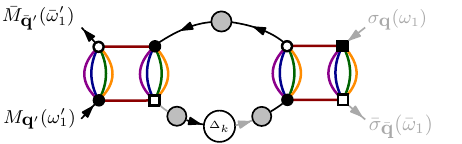}}}\\
&=\left(\vcenter{\hbox{\includegraphics[scale=0.7]{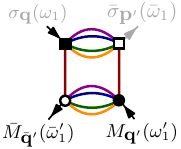}}}+\vcenter{\hbox{\includegraphics[scale=0.7]{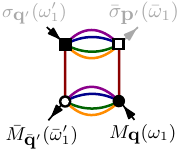}}}\right)\delta_{\bm{p}\bar{\bm{q}}}-\left(\vcenter{\hbox{\includegraphics[scale=0.7]{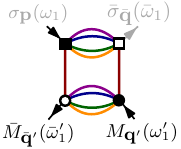}}}+\vcenter{\hbox{\includegraphics[scale=0.7]{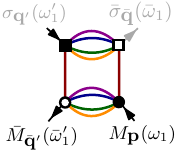}}}\right)\delta_{\bm{p}^\prime \bm{q}}\,,
\end{align}\label{relationWard001}

with the same momentum configuration as before. Let us now translate these relations into explicit equations. Starting from \eqref{relationWard00}, one part reduces to the expression already obtained in the equilibrium dynamics, while accounting for the modifications introduced by the bigamous operators. For the monogamous contribution to $\lambda^\prime$, we find:
\begin{equation}
\lambda^\prime\vert_{\text{mono}}=-\kappa \mathcal{L}+8 \lambda^2 \mathcal{L}_3\,,
\end{equation}
where, as before:
\begin{equation}
\mathcal{L}_3:= \mathcal{L}_3^{(\infty)}+ \mathcal{L}_3^{(0)}\,,
\end{equation}
with:
\begin{equation}
\mathcal{L}^{(0)}_3(k):= k^2 Z(k)\sum_{\bm p \in \mathbb{Z}^4} \frac{\partial }{\partial p_1^2}r_k(\bm p)  \frac{1+\Delta+\Delta^\prime \bm p^2}{(Z \bm p^2+m^2+Z k^2 r_k(\bm p^2))^3}\,.\label{L0}
\end{equation}
As before, the UV counterpart $\mathcal{L}_3^{(\infty)}$ involves momentum integral from deep UV to deep IR, 
\begin{equation}
\mathcal{L}_3^{(\infty)}\propto i Z_{\infty}\int d\omega \sum_{\bm p \in \mathbb{Z}^{4}} G_{k,\bar{M} M}(\bm p,\omega) G_{k,\bar{\sigma} M}(\bm p,\omega)\times \left( 2G_{k,\bar{\sigma} M}(\bm p,\omega)+G_{k,\bar{\sigma} M}(\bm p,-\omega)\right)\bigg\vert_{p_i=0}\,.
\end{equation}
But, here, the integral is superficially convergent, and, following \cite{Lahoche:2018oeo}, in the continuum limit $\Lambda\to \infty$, $\mathcal{L}_3^{(\infty)}$ must goes to zero because of $Z_\infty$. The remaining integral can be easily computed, and we get:
\begin{equation}
\mathcal{L}^{(0)}_3(k)= -\frac{\pi^2}{2k^2Z^2}\frac{1+\bar{\Delta}+\frac{2}{3}\bar{\Delta}^\prime}{(1+\bar{m}^2)^3}\,.
\end{equation}

Finally, because of \eqref{Wardidentity01}, we must have $\mathcal{L} \approx - Z/2 \lambda(k)$, and we obtain the good enough approximation:

\begin{equation}
\lambda^\prime\vert_{\text{mono}}=\frac{\kappa Z}{2 \lambda} -8 \lambda^2 \frac{\pi^2}{2k^2Z^2}\frac{1+\bar{\Delta}+\frac{2}{3}\bar{\Delta}^\prime}{(1+\bar{m}^2)^3}\,,
\end{equation}
which agree with the results of \cite{Lahoche_2020b}, setting $\Delta=\Delta^\prime=0$, as expected. Using dimensionless couplings, we get:
\begin{equation}
\boxed{\bar{\lambda}^\prime\vert_{\text{mono}}=\frac{\bar{\kappa}}{2 \bar{\lambda}} -4\bar{\lambda}^2 \pi^2\frac{1+\bar{\Delta}+\frac{2}{3}\bar{\Delta}^\prime}{(1+\bar{m}^2)^3}\,.}
\end{equation}
The contribution of the bigamous interactions, $\lambda^\prime\vert_{\text{big}}$, requires more careful treatment, since the inner loop is logarithmically divergent and the same trick as before can no longer be applied.
More precisely, the integral typically involves a product of the form $G_{k,\bar{\sigma} M}(-\omega), G_{k,\sigma \bar{M}}(-\omega), G_{k,\sigma \bar{M}}(-\omega)$. A careful analysis of the pole structure shows that this integral closely resembles the one appearing in the definition of $\mathcal{L}$ (\eqref{Linfty}). In both cases, the relevant expression reduces to an integral of the type:
\begin{equation}
\int dy\, \frac{1}{i y +M}\frac{1}{(-i y +M)^2} \sim \frac{1}{M^2}\,.
\end{equation}
Moreover, the equilibrium relation $G_{k,\bar{M} M} = -, G_{k,\bar{\sigma} M}, G_{k,\bar{M} \sigma}$ is expected to remain valid in the deep UV, since the wave-function renormalization of the response field exhibits a vanishing flow in the symmetric phase. The IR instabilities $\Delta$ and $\Delta^\prime$ only arise at later stages, in a regime where the integral is no longer divergent and which collapses into the continuum limit once multiplied by $Z_\infty$. Concretely, we obtain:
\begin{equation}
\lambda^\prime\vert_{\text{big}} =-K \lambda \lambda_2 \mathcal{A}\,,
\end{equation}
where $K$ is a numerical factor that will be computed later (the minus sign appears here due to the conventions for $\lambda_2$, which arises with an additional factor $-i$ compared to $\lambda$), and:
\begin{equation}
\mathcal{A}:=i \int d\omega  \sum_{\bm p \in \mathbb{Z}^{4}} \, \left(Z_\infty+k^2 Z\frac{\partial }{\partial p_1^2}r_k(\bm p)  \right) G_{k,\bar{\sigma} M}(-\omega) G_{k,\sigma \bar{M}}(-\omega) G_{k,\sigma \bar{M}}(-\omega)\,.\label{defA}
\end{equation}
The term proportional to the derivative of the regulator can be readily computed using the regulator, as done previously. We denote it by $\mathcal{A}_0$, and obtain:

\begin{equation}
\mathcal{A}_0=\frac{\pi}{4}\frac{\pi^2}{2Z}\frac{1+\bar{\Delta}+\frac{2}{3}\bar{\Delta}^\prime}{(1+\bar{m}^2)^2}\equiv -\frac{\pi}{4} \mathcal{L}^{(0)}\,.
\end{equation}
Let us focus on the first term of \eqref{defA}. Because what we stated previously, we have:
\begin{equation}
\mathcal{A}^{(\infty)}\approx -iZ_\infty \int d\omega  \sum_{\bm p \in \mathbb{Z}^{4}}G_{k,\bar{\sigma} M}(\omega) G_{k,M \bar{M}}(-\omega) =-\frac{\pi}{4} \mathcal{L}^{(\infty)}\,,
\end{equation}
 and because of \eqref{Wardidentity01}, we have also:
 
\begin{equation}
\mathcal{A}^{(\infty)}\approx\frac{\pi}{4}\left(\frac{Z}{\lambda}+\mathcal{L}^{(0)}\right)\,.
\end{equation}
All that remains is to compute the factor $K$, and it is straightforward to see that, within the chosen momentum configuration, $K = \frac{4 \times 4}{\pi}$. Finally:
\begin{equation}
\boxed{\bar{\lambda}^\prime=\frac{\bar{\kappa}}{2 \bar{\lambda}} -4 \bar{\lambda}^2 \pi^2\frac{1+\bar{\Delta}+\frac{2}{3}\bar{\Delta}^\prime}{(1+\bar{m}^2)^3}-2\bar{\lambda}_2\,.}\label{lambdaprimeF}
\end{equation}
The final piece, $\lambda^\prime_2$, can be obtained from \eqref{relationWard001}, giving us all the ingredients required for the computation. We then find:
\begin{equation}
\lambda^\prime_2=-\kappa_2 \mathcal{L}+4 \lambda \lambda_2 \mathcal{L}_3+\frac{16}{\pi}\lambda_2^2 \mathcal{A}\,.
\end{equation}
Because of the previous result, we obtain finally:
\begin{equation}
 \boxed{\lambda^\prime_2=\frac{\bar{\kappa}_2}{2 \bar{\lambda}}-2\bar{\lambda}\bar{\lambda}_2 \pi^2\frac{1+\bar{\Delta}+\frac{2}{3}\bar{\Delta}^\prime}{(1+\bar{m}^2)^3}+\frac{4\bar{\lambda}^2_2}{\bar{\lambda}}\,.}\label{xprimeF}
\end{equation}
This last relation closes the hierarchy, up to the replacement of $\kappa$ and $\kappa_2$ by $\kappa_{\text{dyn}}$ and $\kappa_{2\text{dyn}}$, and we will give a numerical analysis of the flow equations in section \ref{numan}.

\subsection{Numerical investigations}\label{numan}

In this section, we numerically explore the equations derived above, comparing the semi-perturbative method around the solution provided by the EVE approach with the results obtained via the Ward identities. This section will also retrospectively justify certain assumptions made in the choice of parameterizations.

\paragraph{Perturbative EVE.} The analysis of the flow equations reveals the existence of both isolated fixed points and fixed-point lines, which we now examine separately. Isolated fixed points occur at the following values:

\begin{equation}
P_1=(\bar{m}^2\approx -0.87\,,\bar{\lambda}\approx 0.00017\,,\bar{\Delta} \approx 1.01\,,\bar{\lambda}_2=0,\bar{\Delta}^\prime=0)\,,
\end{equation}
\begin{equation}
P_2=(\bar{m}^2\approx -0.52\,,\bar{\lambda}\approx -0.013\,,\bar{\Delta} \approx -5.05\,,\bar{\lambda}_2=0,\bar{\Delta}^\prime=0)\,.
\end{equation}

\begin{figure}
\begin{center}
\includegraphics[scale=0.55]{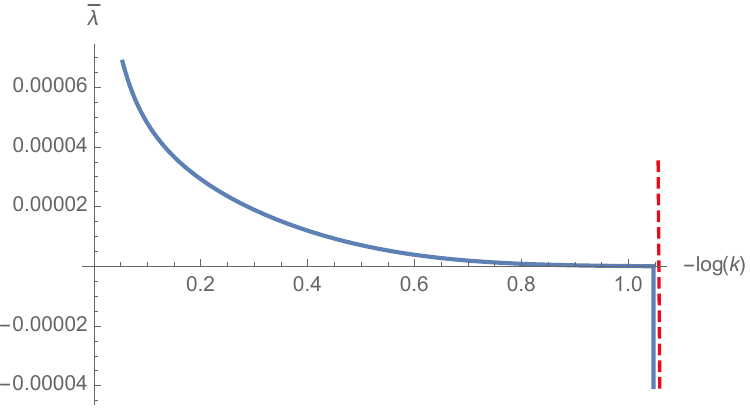}\qquad  \includegraphics[scale=0.55]{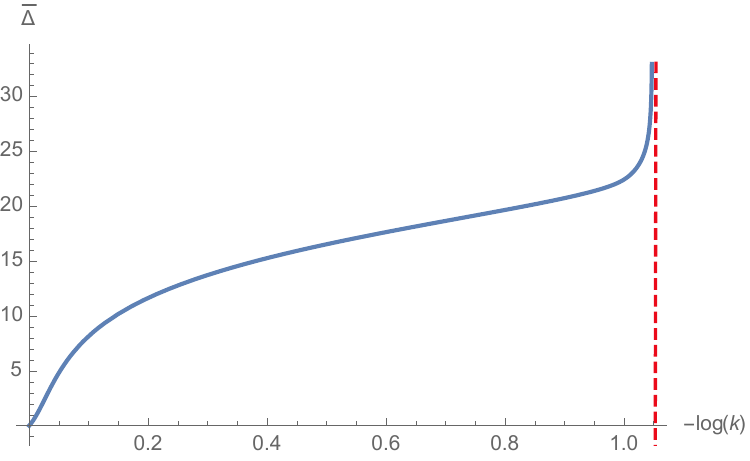}
\end{center}
\caption{Behavior of the RG flow for the initial conditions $\bar{m}^2= -0.87, \bar{\lambda}= 0.00017, \bar{\lambda}_2= 0.01$, the other couplings being zero.}\label{figpro}
\end{figure}

Note that the reliability of these fixed points is rather low, since the magnitudes of the bigamous interactions are large compared to those of the monogamous interactions. The fixed point $P_2$ is likely an artifact of the truncation, as the anomalous dimension $\eta_2 \approx -6.66$ violates the bound \eqref{boundeta}. For $P_1$:

\begin{equation}
\eta_1\approx 0.99\,.
\end{equation}
This is a rather large value for the derivative expansion, once again casting doubt on the reliability of this fixed point (see \cite{Balog_2019}). The calculation of the critical exponents shows that the fixed point $P_1$ possesses only irrelevant directions, along with a single marginal direction:
\begin{equation}
\Theta=(-11.77,-3.08,-2,-11.78,0)\,,
\end{equation}
and therefore essentially corresponds to an IR attractor. Nevertheless, the interpretation that can be drawn is a global instability of the flow within the basin of attraction of the fixed point. Numerical analyses also appear to indicate that finite-scale singularities prevent a perturbation around equilibrium from reaching the fixed point $P_1$, as illustrated in Figure \ref{figpro}. In this context, such behavior seems rather pathological and further supports the view that these fixed points are merely artifacts of our approximations.\\

\begin{figure}
\begin{center}
\includegraphics[scale=0.7]{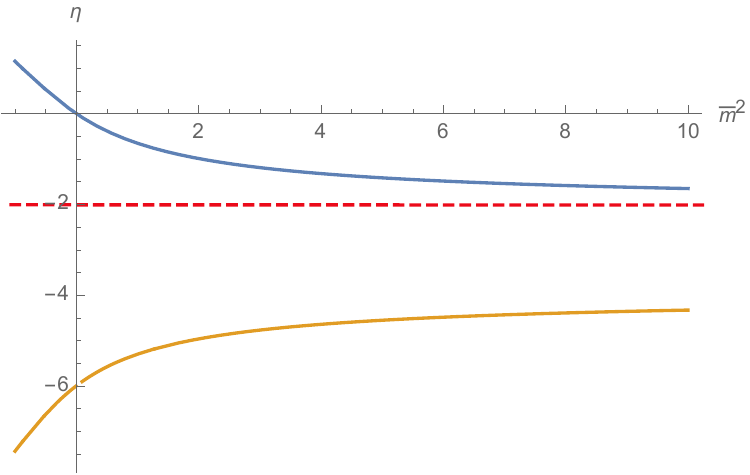}
\end{center}
\caption{Behavior of the anomalous dimension for $L_1$ (in blue) and for $L_2$ (in yellow).}\label{figetaline}
\end{figure}

\begin{figure}
\begin{center}
\includegraphics[scale=0.55]{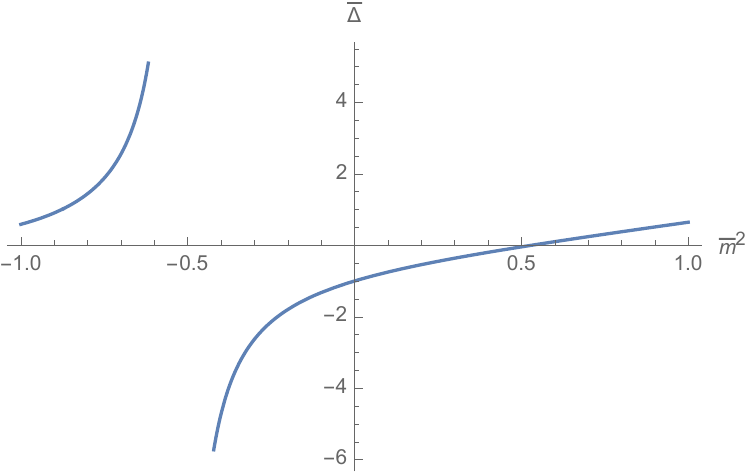}\qquad \includegraphics[scale=0.55]{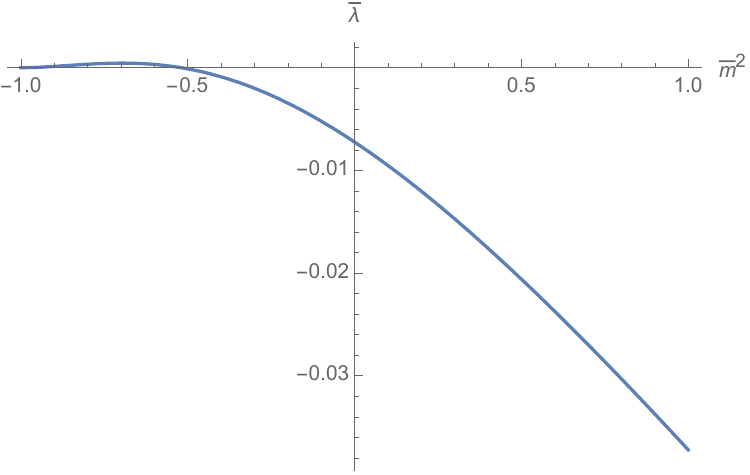}
\end{center}
\caption{Behavior of $\bar{\lambda}$ and $\Delta$ on the fixed point line $L_1$.}\label{L1}
\end{figure}

Now, let us investigate the line of fixed point. The two fixes point lines, $L_1$ and $L_2$ have both the structure:
\begin{equation}
L_1=(\bar{\lambda}=f_i(\bar{m}^2),\bar{\Delta}=g_i(\bar{m}^2))\,,
\end{equation}
with the other couplings set to zero. Figure \ref{figetaline} shows the behavior of the corresponding anomalous dimensions. Only $L_1$ respects the bound \eqref{boundeta} for the anomalous dimension. The behavior of $L_1$ is illustrated in Figure \ref{L1}. However, the only region where $\delta$ is not too large corresponds to a negative value of $\lambda$, around $\bar{m}^2 \approx 0.53$, with $\bar{\lambda} \approx -0.02$ and $\eta \approx -0.40$. Due to the negative sign of $\lambda$, this region lies far from the equilibrium regime (where the path integral must be normalizable and bounded), and is therefore excluded. In conclusion, no reliable fixed point appears to be identifiable using this method.

\begin{figure}
\begin{center}
\includegraphics[scale=0.55]{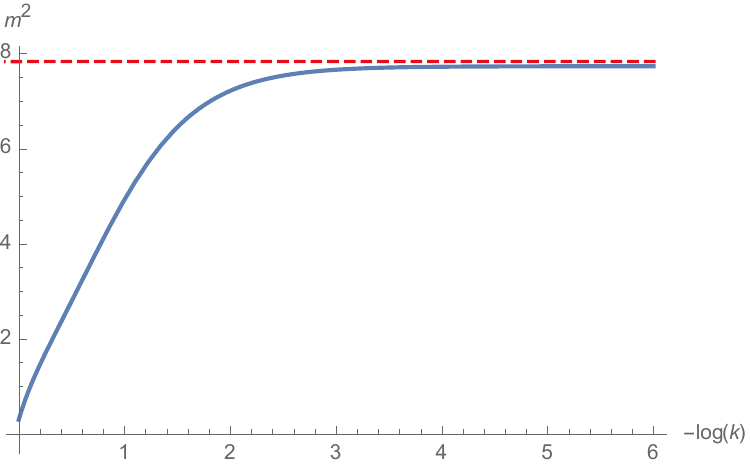}\qquad  \includegraphics[scale=0.55]{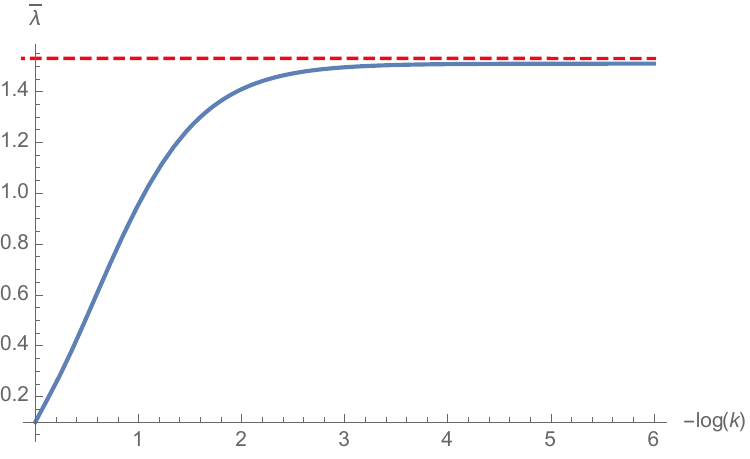}\\
\includegraphics[scale=0.55]{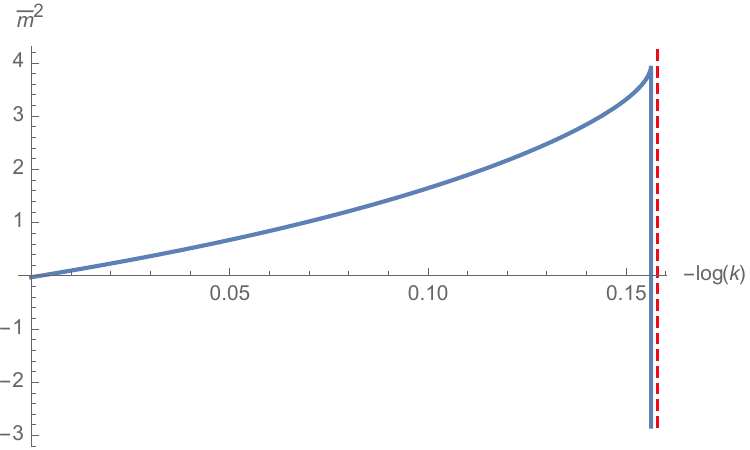}\qquad  \includegraphics[scale=0.55]{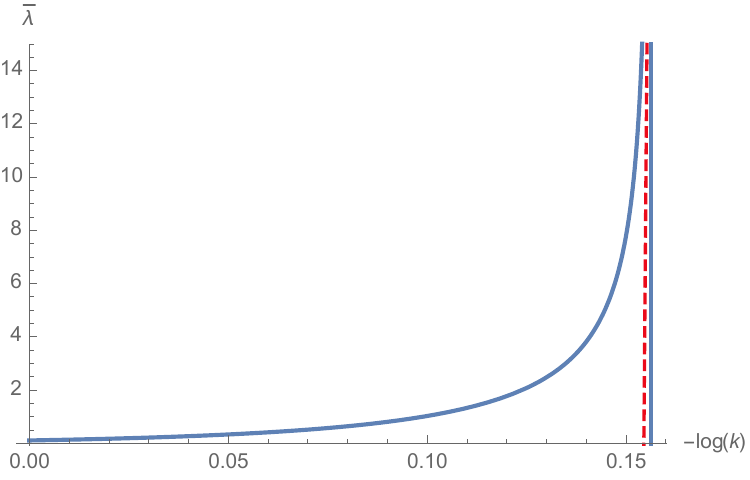}
\end{center}
\caption{On the top: behavior of the RG flow for the initial conditions: $\bar{m}^2(0)=0.3$ and $\bar{\lambda}=0.1$. On the bottom: the behavior of mass and quartic coupling for the initial conditions $\bar{m}^2(0)=-0.03$ and $\bar{\lambda}=0.4$. A finite scale singularity arises for $k \approx \Lambda- 0.86$.}\label{figdivImproved}
\end{figure}

\paragraph{Improved EVE.} We now turn to the investigation of the second method, based on Ward identities. This approach significantly improves upon the EVE method presented in Section \ref{method} by avoiding the approximations associated with the derivative expansion when computing non-perturbative loops in $\Gamma_k^{(6)}$. In this framework, a parameter $\kappa$ is derived directly from the interplay between the flow equations in the melonic approximation and the constraints imposed by the Ward identities, using the derivative expansion only within the momentum window selected by the regulator. As studied in \cite{Lahoche_2020b}, this method shows that no spurious global fixed point arises in the equilibrium theory, leaving the flow unconstrained. Consequently, the phenomenon of finite-scale singularities, predicted by perturbation theory, is recovered, as illustrated in Figure \ref{figdivImproved}. These singularities highlight the complexity of the IR limit and suggest the presence of first-order phase transitions, rather than behavior controlled by an isolated non-perturbative fixed point.\\

Recall that in this case the flow equations are exactly those that we established in section \ref{secWard} by setting $\Delta=\Delta^\prime=\lambda_2=0$. Explicitly\footnote{This expression corrects a little, inconsequential mistake in the arxiv version of the reference \cite{Lahoche_2020b}.}:

\begin{equation}
\eta=-\frac{6 \pi ^2 \bar{\lambda}  \left(\pi ^2 \lambda +2 (1+\bar{m}^2)^3\right)}{\pi ^4 \bar{\lambda} ^2+\pi^2 \bar{\lambda}  (3 \bar{m}^2+5) (1+\bar{m}^2)^2-3 (1+\bar{m}^2)^5}\,,\label{stateeta}
\end{equation}
and:
\begin{equation}
\left\{
\begin{array}{ll}
\dot{\bar{m}}^2&=-(2+\eta)\bar{m}^{2}-\,\frac{10\pi^2\bar{\lambda}}{(1+\bar{m}^{2})^2}\,\left(1+\frac{\eta}{6}\right)\,,\\
\dot{\bar{\lambda}}&=-\eta\bar{\lambda}\, \frac{(1+\bar{m}^2)^2-\pi^2 \bar{\lambda}}{(1+\bar{m}^2)^2}+\frac{2\pi ^2\bar{\lambda}^2}{(1+\bar{m}^2)^3}\dot{\bar{m}}^2\,,
\end{array}
\right.\label{systprime}
\end{equation}
The equation for $\lambda$ is initially fixed by the Ward identity (equation \eqref{wardcont1} without the bigamous interactions). Now, consider the complete set of equations, including all bigamous terms, namely \eqref{betamNew}, \eqref{eqDelta}, \eqref{eqlambda22}, \eqref{eqlambdaKappa}, \eqref{wardcont1}, \eqref{wardcont2}, as well as \eqref{lambdaprimeF} and \eqref{xprimeF}, which are used to compute $\eta$ and $\dot{\bar{\Delta}}^\prime$.\\

Let us first consider the isolated fixed point. Numerical investigations of the resulting flow equations (without imposing any restrictions on the magnitude of the bigamous interactions) indicate that no reliable fixed-point solutions are expected. Indeed, assuming $\eta_* \neq 0$, the conditions $\dot{\bar{\lambda}} = \dot{\bar{\lambda}}_2 = 0$ impose:
\begin{align}
\bar{\lambda}_2+\bar{\lambda} \bar{\Delta}^\prime&=0\,,\\
\bar{\lambda}\pi^2\frac{1+\bar{\Delta}+\frac{2}{3}\bar{\Delta}^\prime}{(1+\bar{m}^2)^2}&=1\,.
\end{align}
Substituting into the flow equations, we determine $\Delta$ from the condition $\dot{\Delta}=0$, and then, inserting this into $\dot{\bar{m}}^2$, we find $\dot{\bar{m}}^2 = 4 \bar{m}^2$, which implies $\bar{m}^2 = 0$. Substituting into the remaining flow equations, we observe that the equations become singular, and no fixed point can be obtained in this way. The alternative condition, setting $\eta_* = 0$, leads only to complex solutions. Hence, no reliable fixed point is expected:
\begin{claim}
No reliable global fixed point is expected in the bigamous non-branching melonic sector at the leading order of the derivative expansion.
\end{claim}

\begin{figure}
\begin{center}
\includegraphics[scale=0.55]{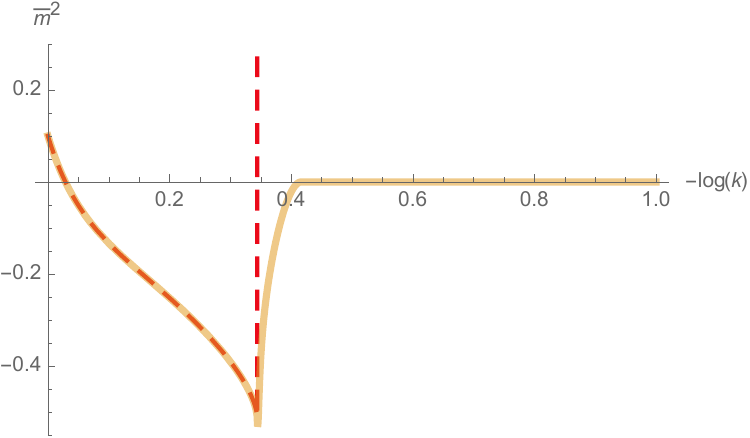}\qquad  \includegraphics[scale=0.55]{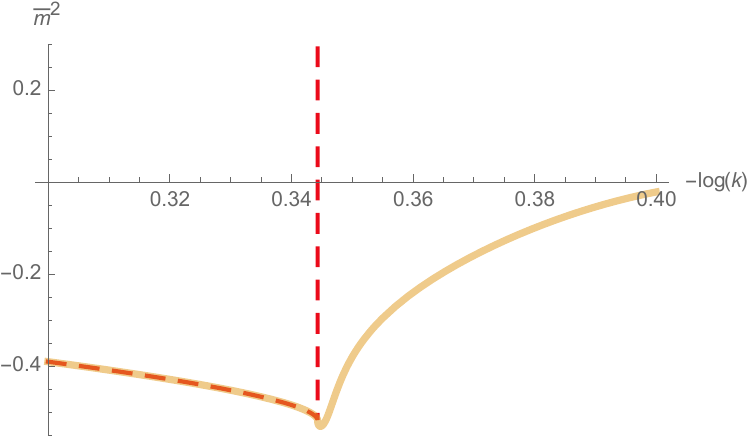}\\
\includegraphics[scale=0.55]{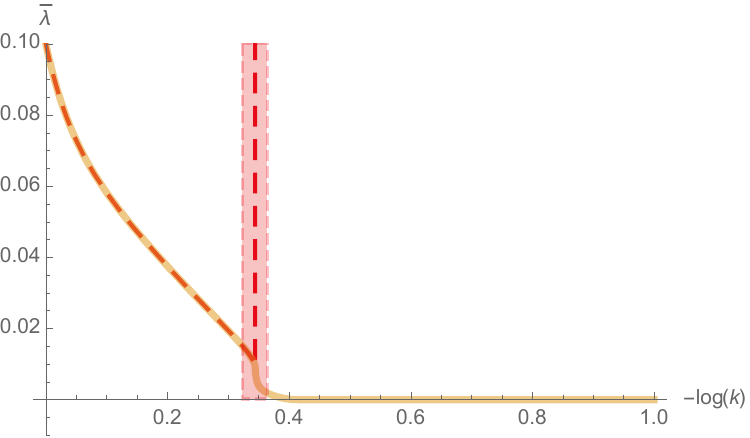}\qquad  \includegraphics[scale=0.55]{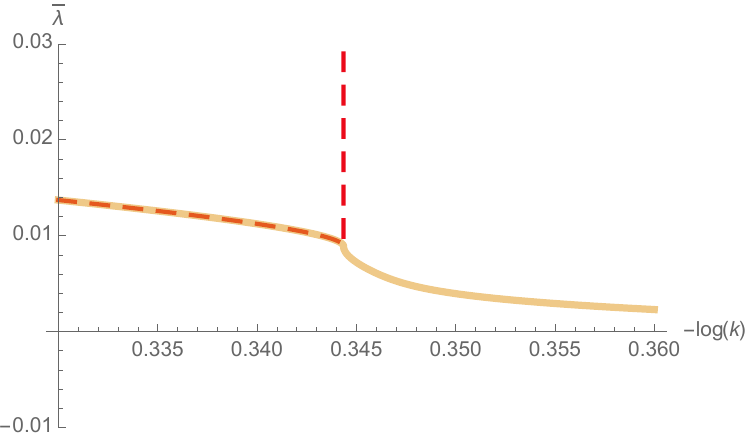}
\end{center}
\caption{Influence of a small deviation of the exact FDT (yellow curve) on a singular trajectory (blue dashed cuve).}\label{figresults1}
\end{figure}

\begin{figure}
\begin{center}
\includegraphics[scale=0.55]{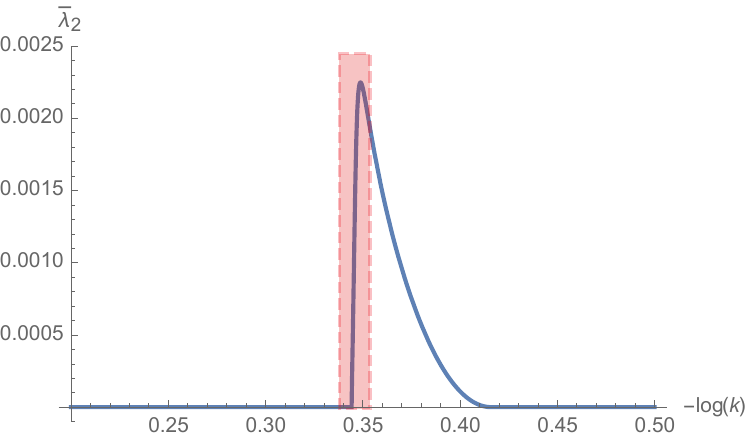}\qquad  \includegraphics[scale=0.55]{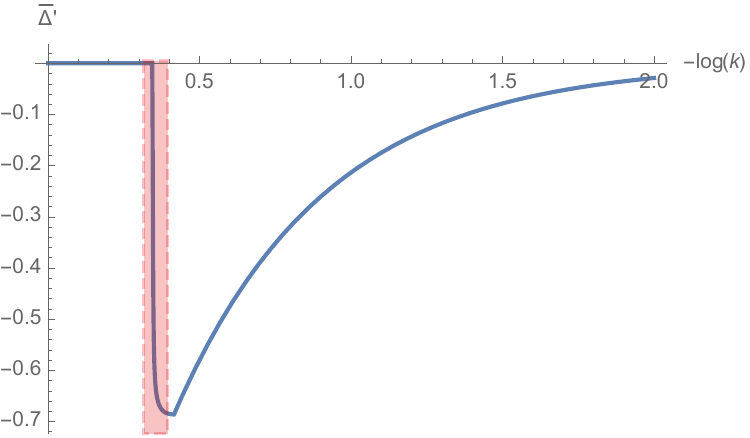}\\
\includegraphics[scale=0.55]{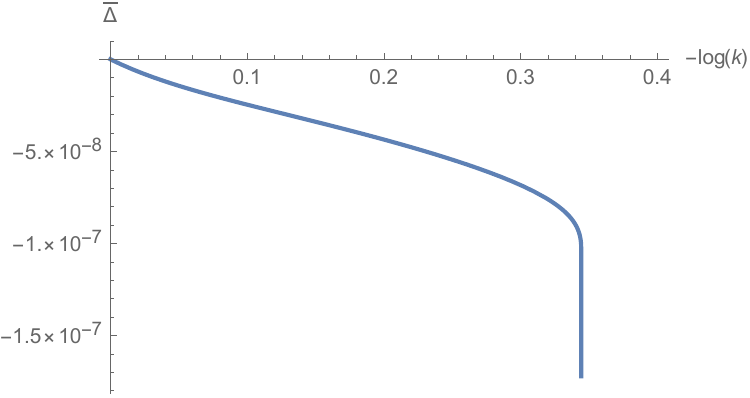}\qquad  \includegraphics[scale=0.55]{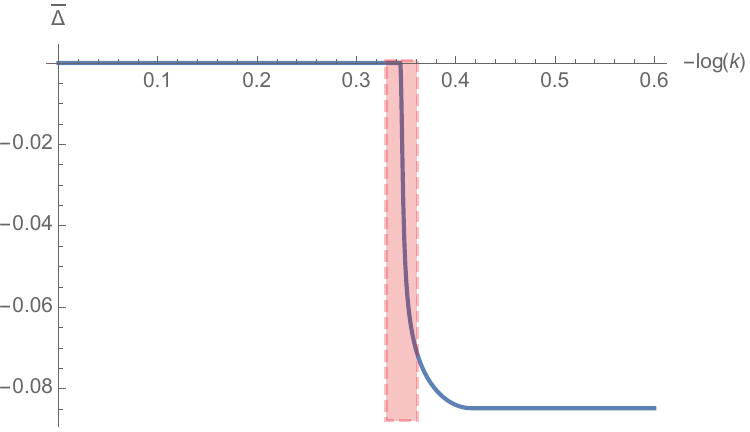}\\
\end{center}
\caption{Behavior of the bigamous interactions along the trajectory.}\label{figresults2}
\end{figure}

\noindent
Let us now turn to the trajectories of the flow. As observed, the flow in the equilibrium dynamics leads to singularities at finite scales. Figures \ref{figresults1}, \ref{figresults1}, and \ref{figresults1} summarize the main results, which we shall now discuss. In Figure \ref{figresults1}, the initial conditions were chosen such that:

\begin{equation}
\bar{m}^2(0)=\bar{\lambda}(0)=0.1\,.\label{initialconditions}
\end{equation}
The red curve shows that the flow exhibits a finite-scale divergence, similar to that seen in Figure \ref{figdivImproved}. The yellow curve is obtained with the same initial conditions as in \eqref{initialconditions}, but with a slight perturbation of the time-reversal symmetry, $\bar{\lambda}_2(0) = 10^{-8}$. In this example, the trajectories remain very close to each other, and, as shown in Figure \ref{figresults2}, the bigamous interactions $\bar{\Delta}$, $\bar{\Delta}^\prime$, and $\bar{\lambda}_2$ remain very small. In other words, the system behaves almost identically to the equilibrium dynamics.

At the point of divergence, however, the bigamous interactions “fly away” (Figure \ref{figresults2}), allowing the system to temporarily avoid the singularity, as indicated by the red region in the figures. Subsequently, the bigamous interactions tend to zero again, except for $\bar{\Delta}$, which remains in a scaling regime at a constant value down to the IR. This suggests that bigamous interactions play a crucial role in shaping the IR behavior in regions where the flow is singular.

These numerical observations support the existence of a discontinuous phase transition to a non-equilibrium regime, which likely corresponds to the IR reality of the system. Importantly, this interpretation is only possible within the stochastic formalism; equilibrium field theory cannot assign a meaningful interpretation to finite-scale singularities. Singularities indicate that equilibrium states are non-normalizable, meaning the system cannot be at equilibrium in regions exhibiting a singularity.

This behavior is not unique; many other numerically identified regions in phase space display similar patterns. Although it would be possible to map these regions, this is beyond the scope of the present work. Finally, not all singular trajectories are resolved by this process, suggesting that other interactions may be relevant. This is the case, for example, for the singular trajectory in Figure \ref{figdivImproved}. Recall that our truncation was chosen for simplicity and primarily to illustrate, beyond methodology, a clue motivated by analytical results obtained in related contexts \cite{lahoche2024functional, kpera2023stochastic}.\\

Let us point out that these results are quite similar to those obtained for example in the recent series \cite{lahoche2025large,achitouv2025constructing,achitouv2024time,natta2024ward,lahoche2024quantum}, where operators forgotten by perturbation theory play a similar role in the construction of the IR limit, avoiding finite-scale singularities. In this case, the argument was still supported by the choice of a $2PI$ formalism, allowing the phenomenon to be interpreted directly as a phase transition. In the present case, this interpretation is indirect, and based on an analogy concerning the behavior of the flow. One could nevertheless hope that a $4PI$ formalism gives similar results, its implementation being however more complex due to the renormalization of the UV divergences. We will return to this in a later work. \\

\begin{figure}
\begin{center}
\includegraphics[scale=0.55]{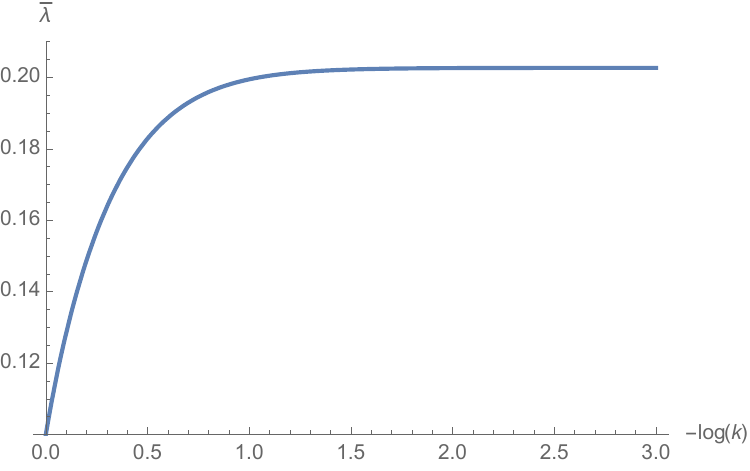}\qquad  \includegraphics[scale=0.55]{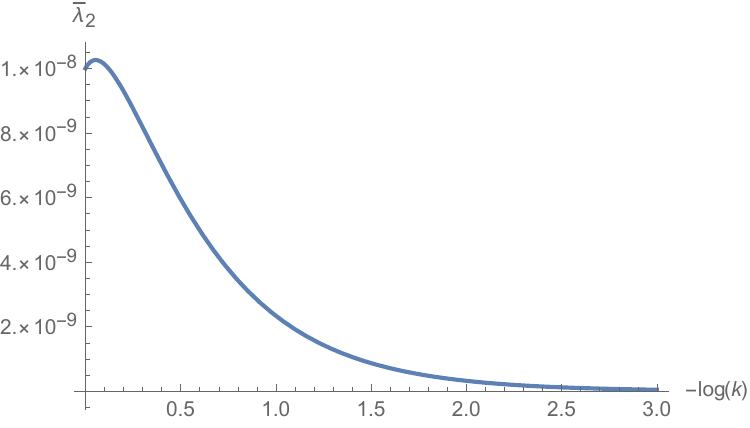}\\
\includegraphics[scale=0.55]{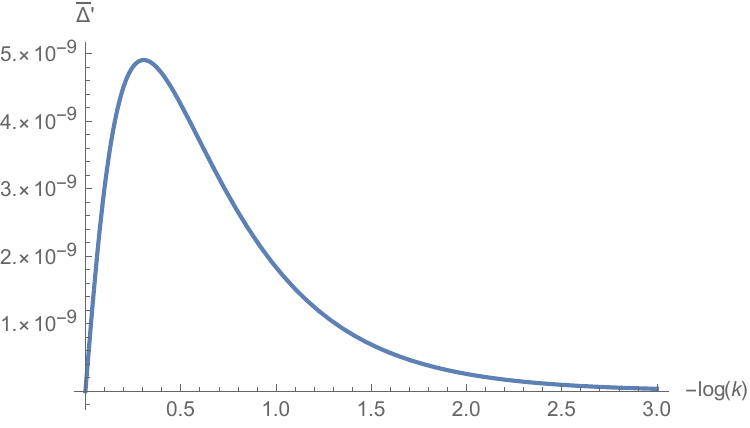}\qquad  \includegraphics[scale=0.55]{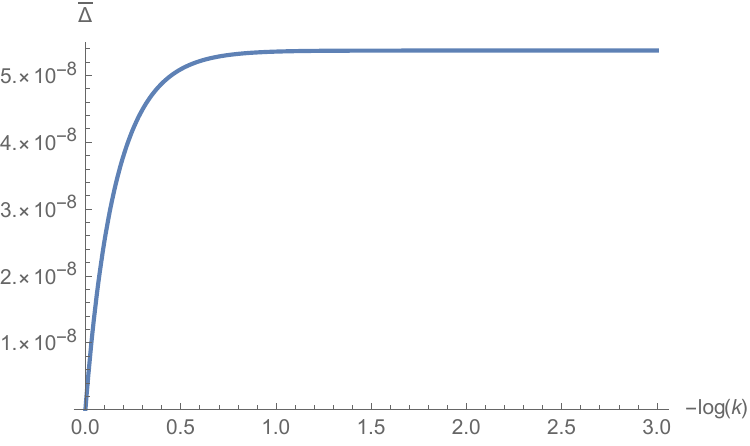}
\end{center}
\caption{Behavior of the RG flow around a convergent trajectory, including a small deviation from FDT.}\label{figresults3}
\end{figure}

Finally, let us mention the case of trajectories without any singularity. Figure \ref{figresults3} illustrates what can be expected in general, for the initial conditions:

\begin{equation}
\bar{m}^2(0)=0.9\,,\qquad \bar{\lambda}(0)=0.1\,.\label{initialconditions2}
\end{equation}
With these initial conditions, the flow converges in the absence of bigamous interactions. Introducing a small perturbation, $\bar{\lambda}_2(0) = 10^{-8}$, we observe that the system remains arbitrarily close to the equilibrium regime throughout the flow, as summarized in Figure \ref{figresults3}. The bigamous interactions tend to zero, except for $\Delta$, which stabilizes at a value of the order of the initial perturbation. This provides a measure of the asymptotic deviation from equilibrium, which remains essentially constant during the flow, illustrating the stability of the equilibrium regime.

\subsection{Time-reversal symmetry}

To conclude, we turn to a specific study of the stability of the time-reversal hypothesis. It should be noted that symmetry breaking is expected even in elementary models. This has been explicitly observed for tensor theories in \cite{kpera2023stochastic}, although the result is quite general and does not rely on the tensorial nature of the field, for instance, a simple Langevin-type model describing the response of a particle in a non-stationary bath exhibits similar behavior \cite{verley2011modified}. \\

The parameterization we consider is inspired by our recent work on similar models: in \cite{lahoche2024functional} for the $p=2$ soft spin dynamics, and in \cite{lahoche2023functional, lahoche2025large, achitouv2024time, achitouv2025constructing} for related issues in the FRG context. We focus on the following truncation for the kinetic effective average action:
 \begin{equation}
\Gamma^{(2)}_k(\vec{p},\omega,\omega^{\prime}) = \begin{pmatrix}
A& B(\bm{p},\omega) \\
B(\bm{p},-\omega)& 0
\end{pmatrix}\delta(\omega-\omega^{\prime})+i\begin{pmatrix}
0& \Delta_k \\
\Delta_k& 0
\end{pmatrix}
\,,\label{truncationTsym}
 \end{equation}
where $\Delta_k$ is a dimensionless constant, independent of $\omega$, and serves as the order parameter for time-translation symmetry. In the absence of further guidance, simplicity provides a natural choice. At first order (neglecting the variable $\bm{p}$ for simplicity), we then obtain:
\begin{equation}
G_{k,\bar{\eta}\Phi}(\omega,\omega^\prime)=B^{-1}(\omega)\delta(\omega-\omega^\prime)-iB^{-1}(\omega)B^{-1}(\omega^\prime) \Delta_k+\mathcal{O}(\Delta_k^2)\,.\label{GLexp}
\end{equation}
Note that this expression is in fact exact, since the next-order term involves the integral $\int \extd \omega, B^{-1}(\omega)$, which vanishes due to the so-called Heteroclyte condition \cite{lahoche2023stochastic}. This condition ensures that interactions not involving the response field cannot be generated\footnote{A simple argument is that $C_{\bar{\sigma} M}(t-t^\prime) \propto \theta(t-t^\prime)$, and the Itô prescription imposes $\theta(0) = 0$.}, in particular guaranteeing that the response field does not propagate—see also \cite{canet2011general}. Importantly, this analysis does not require extending the truncation beyond the kinetic action, as we focus on sufficiently small deviations from equilibrium dynamics. \\

The flow equations for $\lambda$ and $\bar{m}^2$ have, graphically, the same structure as in the equilibrium dynamics. We only need to add to the terms already calculated in Section \ref{method} the corrections arising from the modifications of the propagator, retaining only those contributions that are local in time. In particular, the contributions to $\Delta$ in the equation for $\dot{\Gamma}^{(2)}_k$ are identified in this way, by keeping in the left-hand side the term proportional to $\delta(t), \delta(t^\prime)$.\\

A direct inspection shows that the $\beta$-functions for $\bar{m}^2$ and $\bar{\lambda}$ are unaffected by the non-local contributions. For instance, the mass tadpole in the flow equation for $\bar{m}^2$ typically involves the product $G_{k,\bar{\sigma} M} G_{k,\bar{M} M}$. The order-$\Delta$ correction involves a term of the form $\Delta G_{k,\bar{\sigma} M}(\omega) G_{k,\bar{M} M}(\omega) G_{k,\bar{\sigma} M}(\omega^\prime) \delta(\omega^\prime - \omega - \Omega)$, where $\Omega$ denotes the sum of external frequencies, and the propagators in the product are taken at zeroth order in $\Delta$. The Dirac delta from the interaction is canceled by the integration over $\omega^\prime$, and the expansion to zeroth order in $\Omega$ produces a contribution only to the flow of $\Delta$. The same reasoning applies to all loops: locality is destroyed by any non-local insertion. Consequently, $\dot{\bar{m}}^2$ and $\dot{\lambda}$ are again given by their equilibrium dynamics expressions:

\begin{equation}
\dot{\bar{m}}^2=-(2+\eta)\bar{m}^{2}-\frac{10 \pi^2 \bar{\lambda}}{(1+\bar{m}^{2})^2}\,\left(1+\frac{\eta}{6}\right)\,,\label{eqmasse2}
\end{equation}

\begin{align}
\dot{\bar{\lambda}}=-2\eta \bar{\lambda}-3 \pi^2 \bar{\kappa}_2\,\frac{1}{(1+\bar{m}^{2})^2}\,\left(1+\frac{\eta}{6}\right)+4\pi^2\bar{\lambda}^2 \,\frac{1}{(1+\bar{m}^{2})^3}\,\left(1+\frac{\eta}{6}\right)\,.\label{eqlambdaKappa2}
\end{align}

The expression for $\eta$ is also the same as in the equilibrium regime, explicitly:

\begin{equation}
\eta=-\bar{\lambda}^{\prime}L_{21}(0)-\bar{\lambda} \frac{d}{d p_1^2} L_{21}(x_1)\big\vert_{x_1=0}\,,\label{eqeta}
\end{equation}
explicitly:
\begin{equation}
\eta=-2\bar{\lambda}^\prime \pi^2 \,\frac{1}{(1+\bar{m}^{2})^2}\,\left(1+\frac{\eta}{6}\right)+\frac{\bar{\lambda} \pi^2}{(1+\bar{m}^{2})^2} \left(4+\eta\right)\,,
\end{equation}
which can be solved as:
\begin{equation}
\eta=2\pi^2\frac{2\bar{\lambda}-\bar{\lambda}^\prime}{(1+\bar{m}^{2})^2+\bar{\lambda}^\prime \pi^2/3- \bar{\lambda}\pi^2}\,.
\end{equation}

\begin{remark}
As pointed out in \cite{Lahoche_2020b}, the presence of $+\bar{\lambda}^\prime \pi^2/3$ in the denominator avoid the singularity in the region $\bar{m}^2>-1$, and then, extend maximally the region where the symmetric phase can be considered comparing with standard vertex expansion.
\end{remark}

Because of the previous discussion, the explicit expression for $\dot{\Delta}$ can be deduced from the melonic contributions in the tadpole diagrams of \eqref{flowmassdiag}. At the order $1$ in $\Delta$, we easily get:
\begin{equation}
\dot{\Delta}=-\left(\pi^2 \bar{\lambda}\,\frac{1+\frac{\eta}{6}}{(1+\bar{m}^{2})^2}\,\right) \Delta\,.
\end{equation}
Everywhere in the symmetric phase, in the region $\lambda>0$ where stability of the expansion is expected, $\Delta$ increases toward IR scale.

\section{Conclusion}

In this paper, we have developed a general formalism for the nonperturbative renormalization group of a stochastically quantized TGFT in the symmetric phase. The proposed framework, building on \cite{Lahoche:2018oeo,lahoche2023stochastic} and inspired by the recent work of the same authors on disordered matrix fields \cite{lahoche2024functional}, goes beyond the standard vertex expansion, which often leads to poorly controlled approximations \cite{Carrozza_2017,Lahoche_2017bb}. By restricting ourselves to the regime of non-branched melons, we showed that the hierarchy of flow equations can be closed around the just-renormalizable sector, as determined by the power counting of the equilibrium theory. The method combines Ward identities, made non-trivial due to the form of the propagator, with Schwinger–Dyson equations in the melonic limit. This reduces the problem to a small set of equations, whose approximations can be benchmarked at equilibrium against analytic results, for instance in the large-$d$ limit \cite{lahoche2021large}.

We investigated two types of perturbations, and in both cases we observed an instability of the equilibrium dynamics. While this outcome is limited to a toy model with little direct physical relevance, it could nonetheless extend to other models. Studying such cases within this general methodology will be the subject of future work.

Finally, let us mention some possible extensions. First, the formalism could be generalized to branched sectors, following the method outlined in \cite{Lahoche:2018oeo}. The truncation of the two-point function remains a limitation of the present approach, and a systematic study of its impact would provide a useful criterion of robustness. More fundamentally, the intrinsic constraints of the method, its reliance on the non-branching sector and its focus on the UV regime, appear more challenging to overcome. We plan to return to these issues in future investigations.

\printbibliography[heading=bibintoc]
\end{document}